\documentclass[conference]{IEEEtran}
\usepackage{cite}
\ifCLASSINFOpdf
  \usepackage[pdftex]{graphicx}
\else
   \usepackage[dvips]{graphicx}
\fi
\usepackage{amsmath}
\usepackage{amssymb}
\usepackage{adjustbox}
\usepackage{algorithm}
\usepackage{algorithmic}
\usepackage{array}
\usepackage{url}
\usepackage{tabularx}
\usepackage{booktabs}
\usepackage{epsfig}
\usepackage[hidelinks=true]{hyperref}
\usepackage{subfigure}
\usepackage{tikz}

\begin{document}
\title{FRaZ: A Generic High-Fidelity Fixed-Ratio Lossy Compression Framework for Scientific Floating-point Data}

\author{\IEEEauthorblockN{Robert Underwood\IEEEauthorrefmark{1}\IEEEauthorrefmark{2},
Sheng Di\IEEEauthorrefmark{2},
Jon C. Calhoun\IEEEauthorrefmark{3}, 
Franck Cappello\IEEEauthorrefmark{2}}
\IEEEauthorblockA{\IEEEauthorrefmark{1}School of Computing Clemson University, Clemson, SC 29634}
\IEEEauthorblockA{\IEEEauthorrefmark{2}Argonne National Laboratory, Lemont, IL 60439}
\IEEEauthorblockA{\IEEEauthorrefmark{3}Holcolmbe Department of Electrical and Computing Engineering\\ Clemson University, Clemson, SC 29634}}
\maketitle

\begin{abstract}
With ever-increasing volumes of scientific floating-point data being produced by high-performance computing applications, significantly reducing scientific floating-point data size is critical, and error-controlled lossy compressors have been developed for years.
None of the existing scientific floating-point lossy data compressors, however, support effective fixed-ratio lossy compression. Yet
fixed-ratio lossy compression for scientific floating-point data not only compresses to the requested ratio but also respects a user-specified error bound with higher fidelity.
In this paper, we present FRaZ: a generic fixed-ratio lossy compression framework respecting user-specified error constraints.
The contribution is twofold.
(1) We develop an efficient iterative approach to accurately determine the appropriate error settings for different lossy compressors based on target compression ratios.
(2) We perform a thorough performance and accuracy evaluation for our proposed fixed-ratio compression framework with multiple state-of-the-art error-controlled lossy compressors, using several real-world scientific floating-point datasets from different domains.
Experiments show that FRaZ effectively identifies the optimum error setting in the entire error setting space of any given lossy compressor.
While fixed-ratio lossy compression is slower than fixed-error compression, it provides an important new lossy compression technique for users of very large scientific floating-point datasets.
\end{abstract}
\IEEEpeerreviewmaketitle

\section{Introduction}
\label{sec:intro}

Today's scientific research applications produce volumes of data too large to be stored, transferred, and analyzed efficiently because of  limited storage space and potential bottlenecks in I/O systems. Cosmological simulations \cite{hacc,nyx}, for example, may generate more than 20 PB of data when simulating 1 trillion particles over hundreds of snapshots per run. Climate simulations, such as the Community Earth Simulation Model (CESM) \cite{cesm}, may produce hundreds of terabytes of data \cite{foster2017computing} for each run. 

Effective data compression methods have been studied extensively. Since the major scientific floating-point datasets are composed of floating-point values, however, lossless compressors \cite{fpzip,zstd,gzip}
cannot effectively compress such datasets because of high entropy of the mantissa bits. Therefore, error-bounded lossy compressors have been widely studied because they not only significantly reduce the data size but  also prevent data distortion according to a user's specified error bound. Most existing lossy compressors consider the error bound preeminent and endeavor to improve the compression ratio and performance as much as possible subject to the error bound. 

However, many  scientific application users have requirements for the compression ratio. These requirements are determined by multiple factors such as the capacity of the assigned storage space, I/O bandwidth, or desired I/O performance. Hence, these users desire to perform fixed-ratio lossy compression---that is, compressing data based on the required compression ratio instead of only strictly respecting user's error bound. In this case, the lossy compressor needs to adjust the error bound to respect the target user-specified compression ratio, while minimizing the data distortion. The user can also provide additional constraints regarding the data distortion (such as maximum error bound) to guarantee the validity of the results from reconstructed data. While fixed-ratio compression can be obtained by simply truncating the mantissa of the floating-point numbers, this approach may not respect the user's diverse error constraints. With such additional constraints, the lossy compressor should make the compression ratio approach the expected level as closely as possible, while strictly respecting the data distortion constraints. 

In this paper, we propose a generic, efficient fixed-ratio lossy compression framework, FRaZ, that is used to determine the error settings accurately for various error-controlled lossy compressors, given the particular target compression ratio with a specific scientific floating-point dataset. Our design involves two critical optimization strategies. First, we develop a global optimum searching method by leveraging Davis King's global minimum finding algorithm \cite{kingDlibLibraryOptimization2018} to determine the most appropriate error setting based on the given compression ratio and dataset in parallel. Second, our parallel algorithm optimizes the parameter searching performance by splitting the search range into distinct regions, parallelizing on file, and---in the offline case---by time-step.

Constructing a generic, high-fidelity framework for fixed-ratio lossy compression poses many research challenges.
First, as we elaborate in Section~\ref{sec:design-opt}, the relationship between error bounds and compression ratios is not always monotonic because of the use of dictionary encoder phases in some compressors such as SZ.
Second, since we aim to create a generic framework such that more compressors can be included in the future, we cannot utilize properties of the specific compressors we use to optimize performance  as has been done in prior work \cite{xincluster2018, taoOptimizingLossyCompression2019} and must instead treat the compression algorithm as a black box.
This means that our algorithm cannot take advantage of properties induced by block size or expected behavior induced for a particular data distribution.
Third, since we treat the compressors as a black box, we must carefully study how to modify existing algorithms to minimize the calls to the underlying compressors and orchestrate the search in parallel, in order to have a tool that is useful to users while working around the limitations of the various current and potential future compressors.

We perform the evaluation for our framework based on the latest versions of state-of-the-art lossy compressors (including SZ  \cite{sz16,sz17,liangErrorControlledLossyCompression2018},
ZFP \cite{zfp}, and MGARD \cite{mgard}), using well-known real-world scientific floating-point datasets from the public Scientific Data Reduction Benchmark (SDRBench) \cite{sdrb}. We perform the parallel performance evaluation on Argonne's Bebop supercomputer \cite{bebop} with up to 416 cores. Experiments show that our framework can determine the error setting accurately within the user-tolerable errors based on the target compression ratios, with very limited time overhead in real-world cases. 

The remainder of the paper is organized as follows. In Section \ref{sec:background}, we introduce the background of this research regarding various state-of-the-art lossy compressors, and we present several examples about user's requirement on specific compression ratios. In Section \ref{sec:relate}, we compare  our work with the related work from fixed-rate compression,
image processing, and signal processing. In Section \ref{sec:formulation}, we present a formal problem formulation to clarify our research objective. In Section \ref{sec:design-opt}, we describe our design and our performance optimization strategies. In Section \ref{sec:evaluation}, we present the evaluation results.  In Section \ref{sec:conclusion}, we present or conclusions and end with a vision of future work. 

\section{Background}
\label{sec:background}

In this section, we describe the research background, including the existing state-of-the-art error-controlled lossy compressors and fixed-ratio use cases. 

\subsection{Error-Bounded Lossy Compression}
\subsubsection{SZ} SZ has been widely evaluated in the scientific floating-point data compression community \cite{Baker-Climate17, N-body-compression,qcsim-compression, Peter-error-distribution}
, showing that it is one of the best compressors in its class.  

SZ is designed based on a blockwise prediction-based compression model. It splits each dataset into many consecutive non-overlapped blocks (such as 6$\times$6$\times$6 for a 3D dataset) and performs compression in each block. It includes four key steps: 
\begin{itemize}
    \item \textit{Step 1: data prediction.} SZ adopts a hybrid data prediction method (either a 1-layer Lorenzo predictor \cite{lorenzo} or linear regression method) to predict each data point by its neighboring values in the multidimensional space.
    \item \textit{Step 2: linear-scaling quantization.} Each floating-point data value is converted to an integer number in terms of the formula $quantization\_code = \frac{predicted\_value - true\_value}{2\epsilon}$, where $\epsilon$ refers to the user-specified error bound (i.e., linear-scaling quantization).
    \item \textit{Step 3: entropy encoding.} A Huffman encoding algorithm customized for integer code numbers is then applied to the quantization codes generated by Step two. 
    \item \textit{Step 4: dictionary encoder.} A dictionary encoder such as Gzip \cite{gzip} or Zstd \cite{zstd} is used to significantly reduce the Huffman-encoded bytes generated from Step three. 
\end{itemize}

SZ allows one to set an absolute error bound to control the data distortion in the compression.

\subsubsection{ZFP} 

ZFP \cite{zfp} is another outstanding error-controlled lossy compressor and is also broadly assessed and used in the scientific floating-point data compression community. ZFP transforms floating-point data to fixed-point values block by block (block size is 4$\times$4$\times4$ for 3D datasets) and adopts an embedded coding to encode the generated coefficients.

ZFP also provides an absolute error bound to control the data distortion. Although ZFP provides another fixed-rate compression mode, 
allowing users to do the data compression based on a given compression ratio, the compression quality is significantly worse than the absolute error-bound mode. We demonstrate this in Section~\ref{sec:visualquality}.
Thus, how to efficiently fix compression ratio based on absolute error-bound mode is critical to ZFP. 

\subsubsection{MGARD}

MultiGrid Adaptive Reduction of Data (MGARD) \cite{mgard} is an error-controlled lossy compressor supporting multilevel lossy reduction of scientific floating-point data.
MultiGrid is designed based on the theory of multigrid methods \cite{Trottenberg:2000:MUL:374106, JWRuge_KStuben_1987a}.
An important feature of MGARD is providing  guaranteed, computable bounds on the loss incurred by the data reduction.
MGARD provides different types of norms, such as infinity norm and L2 norm, to control the data distortion.
The infinity norm is equivalent to the absolute error bound, and the L2 norm mode can be used to control the mean squared error (MSE) during the lossy compression. 

Although SZ, ZFP, and MGARD provide advanced features to control the distortion of lossy compression, none of them provide high-fidelity fixed-ratio compression.

Accuracy of EBLC is dictated at compression time by selection of an error bound and error bounding type~---~e.g., absolute, relative, number of bits and are selected to minimize impact on quantities of interest in scientific simulations. For use cases that preform data analytics on lossy compressed data, trial-and-error is often used to identify acceptable compression tolerances~\cite{bakerEvaluatingImageQuality2019}. The trial-and-error is often done offline to ensure that the selected error bound is robust for multiple time-steps and does diminish the quality of the analysis. However, if the lossy compressed data is used to advance the simulation the simulation trial-and-error is possible~\cite{IPDPS_kento}, but recent works have explored the relation of compression error to numerical errors present in the simulation and provide strategies on error tolerance selection~\cite{ Peter-error-distribution, CalhounIJHPCA:LossyCompression, errZFP, Pavlo-Cluster2019-lossyCR}. 

\subsection{Fixed-Ratio Use Cases}

In this subsection, we describe several fixed-ratio use cases to demonstrate the practical demands on the fixed-ratio compression by real-world application users.  

The first use case is significant reduction of storage footprint. On the ORNL Summit system, for example, the capacity of the storage space is limited to 50 TB for each project by default. 
Many scientific floating-point simulations (such as the CESM climate simulation and HACC cosmological simulation) may produce hundreds of terabytes of data in each run (or even over 1 PB of data), such that the compression ratio has to be 10:1 or higher to avoid execution crash due to no space being left on storage.
Even if a larger storage allocation is awarded or purchased at considerable financial cost, projects generating extreme volumes often face the need to reduce their storage footprint in order to make room for their next executions. Fixed-rate compression provides the ability to store multiple simulations given a fixed amount of storage but suffers from large inaccuracies in the data which high compression ratios are required (see Figure~\ref{fig:nyx-data-vis-temp}).


The second practical use case explores best-fit lossy compression solutions based on the user's post-analysis requirement (such as visual quality or specific analysis property) by at fixed compressed sizes. None of the existing error-controlled lossy compressors provide the fixed-ratio compression mode, however, and therefore users have to seek the best-fit choice by conducting inefficient trial-and-error strategies with different error settings for each compressor to achieve a target compression ratio. Furthermore, there is no universal model that accurately predicts compression ratio based on compressor configuration for a variety of input data. 

The third practical use case involves the matching of I/O bandwidth constraints and accelerating the I/O performance.
Advanced light-source instruments, such as the Advanced Photon Source and Linac Coherent Light Source (LCLS-II), may generate image data at an extremely high acquisition rate, such that the raw data cannot be stored efficiently for post-analysis because of limited I/O bandwidth.
Specifically, LCLS-II is producing  instrument data with up to 250 GB/s while the corresponding storage bandwidth is only 25 GB/s. Thus the designers of LCLS-II expect to reduce the data size with a compression ratio of 10 or higher \cite{franck-ijhpca19}.
Spring8 \cite{spring8} researchers also indicate that their data could be generated with 2 TB/s, which is expected to be reduced to 200 GB/s after  data compression. 

We note that users often require random-access decompression across time steps, which means that they prefer to be able to decompress the data individually at each time-step because decompressing the whole dataset with all time-steps requires a significant amount of time or is impossible because of memory allocation limits. 

\section{Related Work}
\label{sec:relate}
In the compression community, the similar type of compression is called ``fixed-rate compression,'' where the \textit{rate} here refers to the bit rate, which is defined as the number of bits used to represent one symbol (or data point) on average after compression. The lower the bit rate, the higher the compression ratio. Hence, fixing the bit rate means fixing the compression ratio.
In the remainder of this section, we compare  our work with prior work in these areas.

\begin{figure}[ht] \centering
\hspace{-0mm}
\subfigure[{original raw data}]
{
\includegraphics[scale=.33]{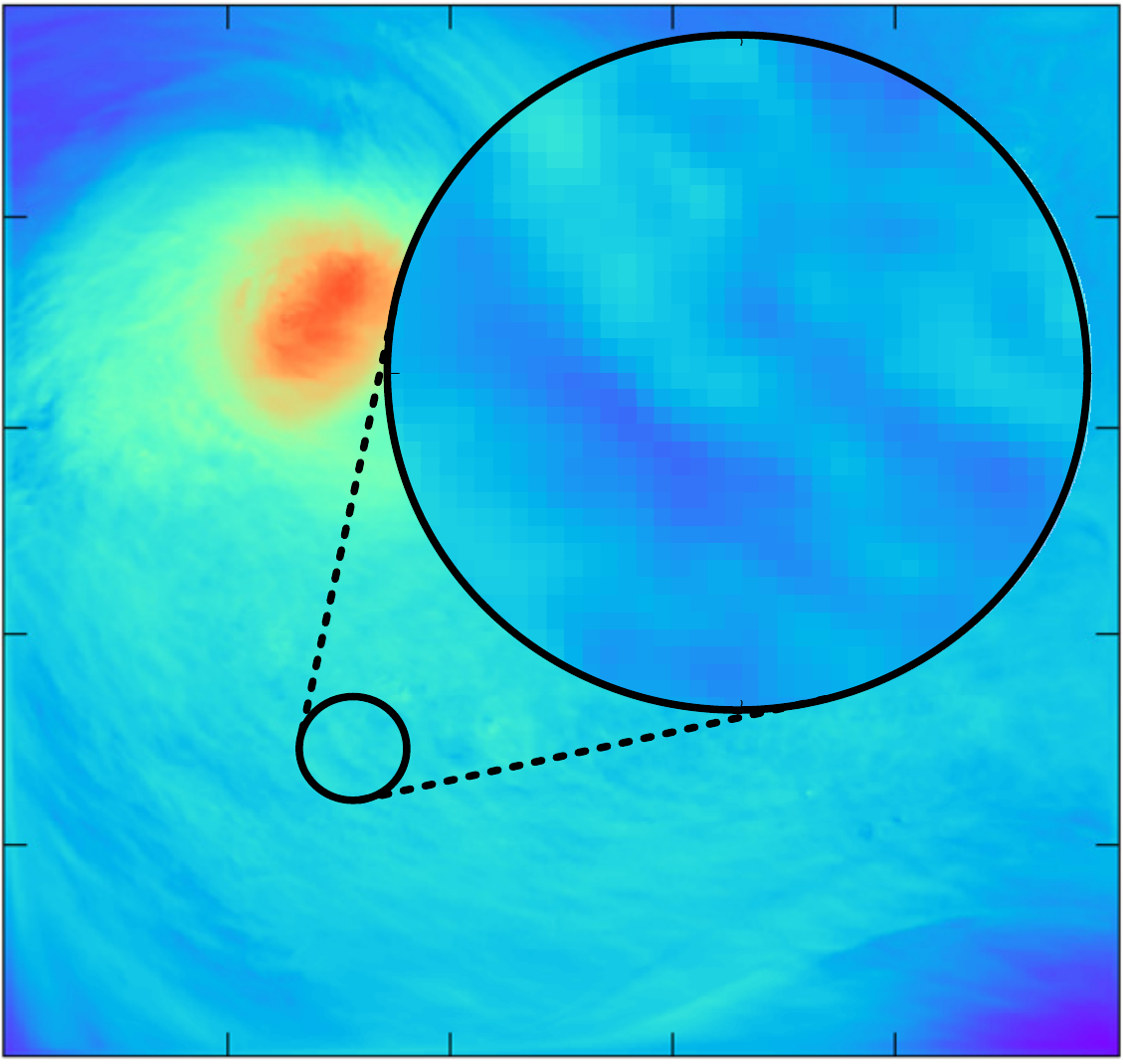}
}
\hspace{-12mm}
\subfigure[{Rate Distortion}]
{
\includegraphics[scale=.43]{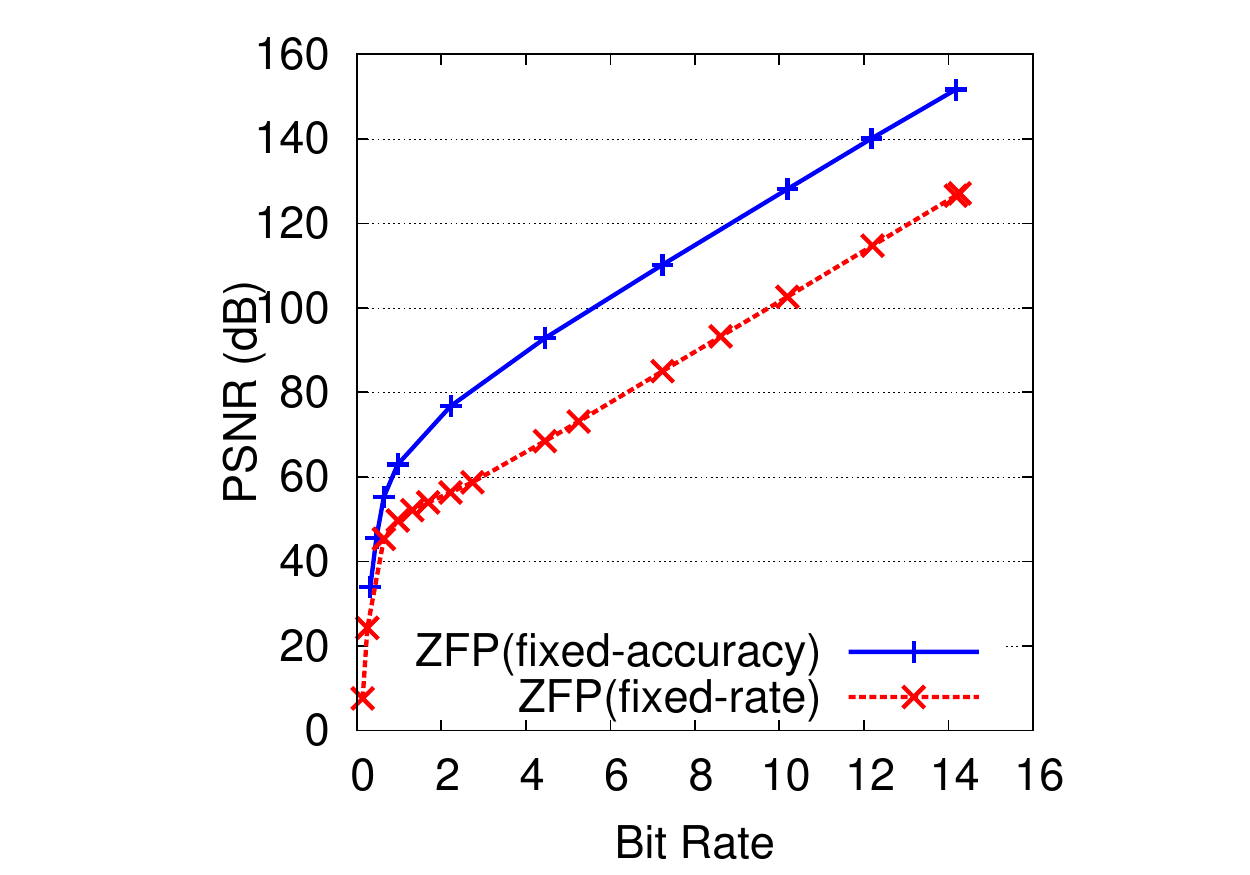}
}
\hspace{-6mm}

\hspace{-7mm}
\subfigure[{ZFP(fixed-accuracy)}]
{
\includegraphics[scale=.33]{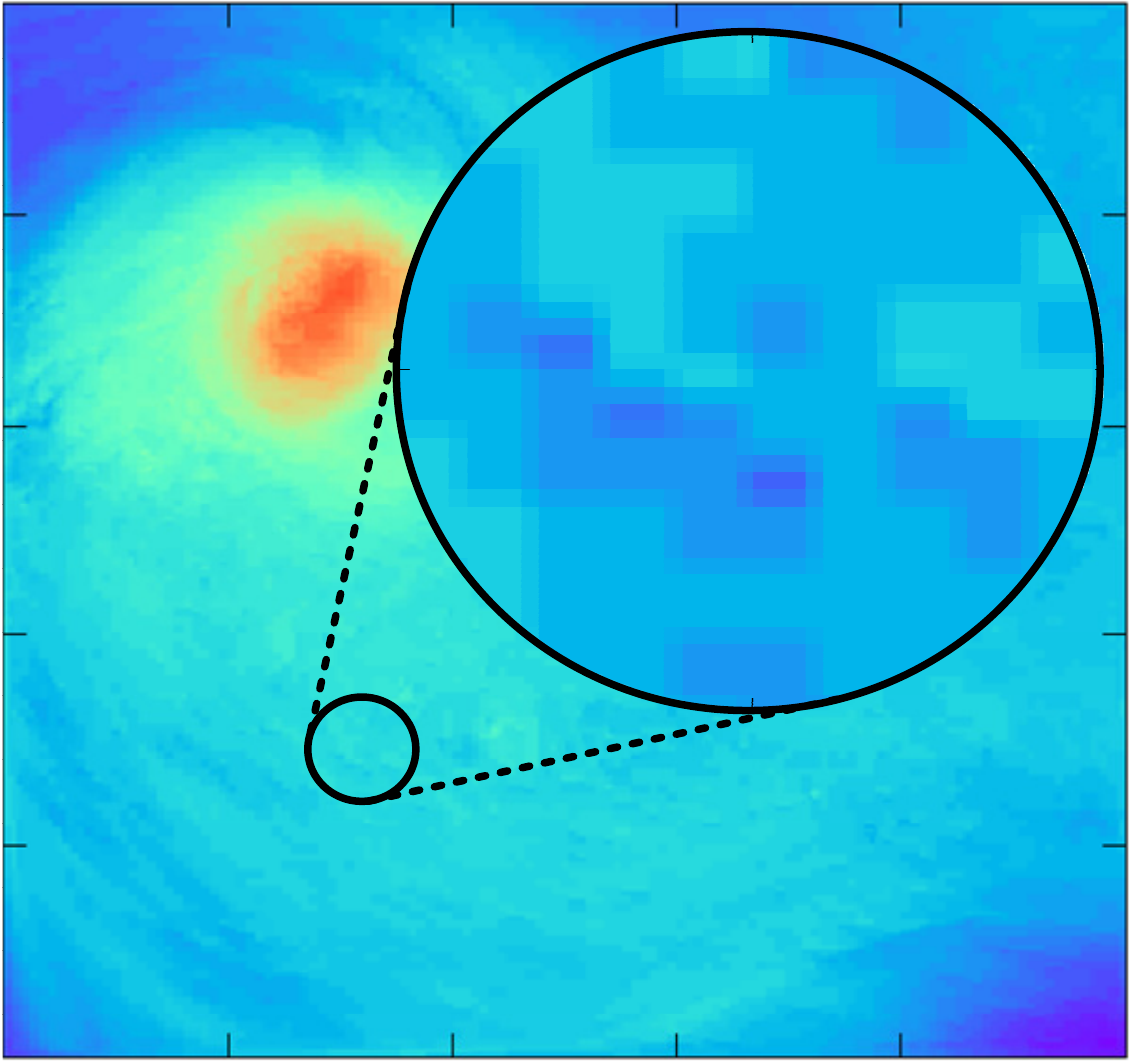}
}
\hspace{-3mm}
\subfigure[{ZFP(fixed-rate)}]
{
\includegraphics[scale=.33]{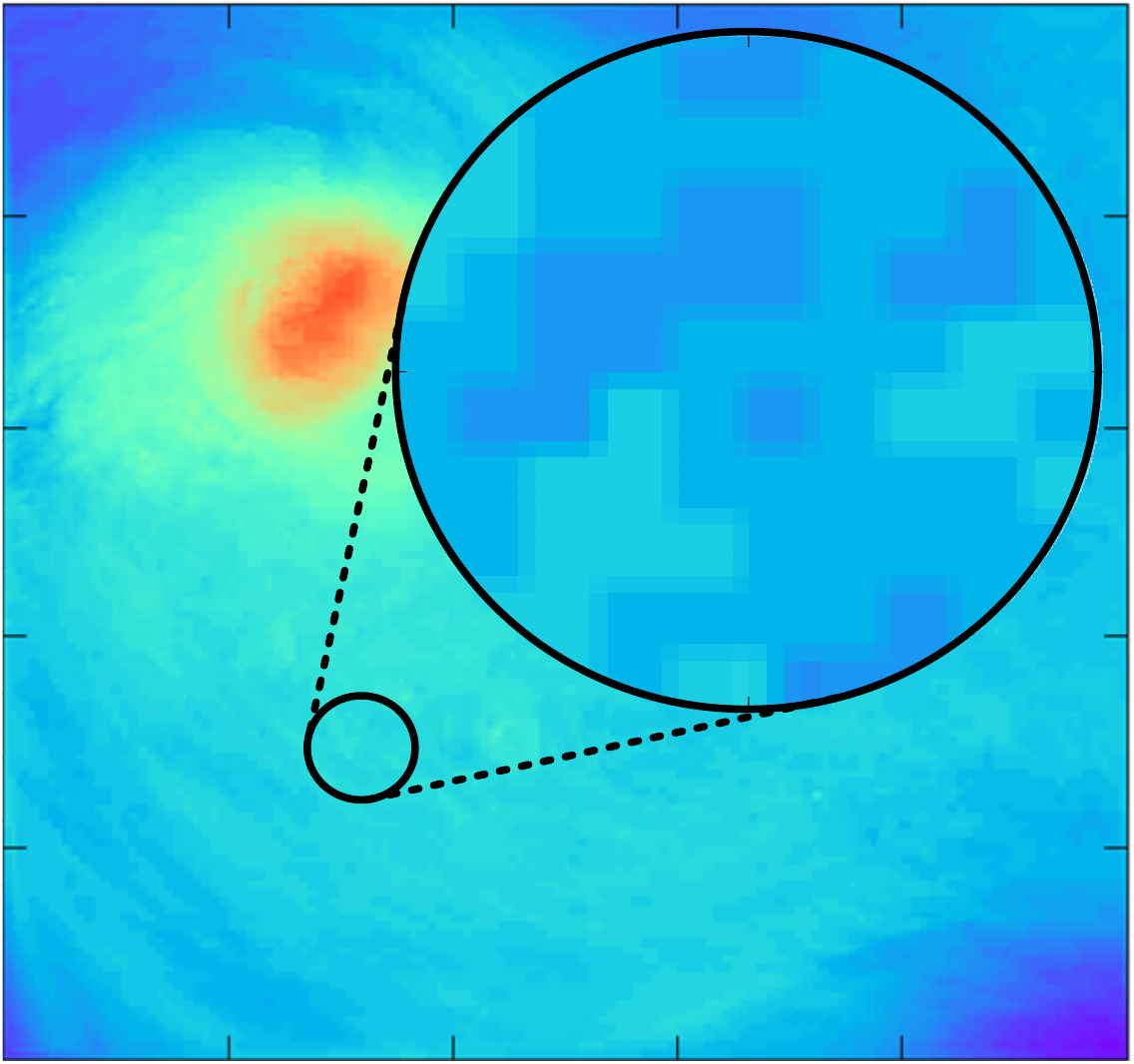}
}
\hspace{-6mm}

\vspace{-2mm}
\caption{Data distortion of ZFP in fixed-accuracy mode and fixed-rate mode (Hurricane TCf field) with CR=50:1; (ZFP.fixed-accuracy: PSNR=55.3, max error=4.2, SSIM=0.94, ACF(error)=0.67; ZFP.fixed-rate: PSNR=45.4, max error=33.7, SSIM=0.94, ACF(error)=0.72)}
\label{fig:hurricane-tcf-zfp-vis}
\end{figure}

In addition to the fixed-accuracy modes (i.e., accuracy and precision error-bounding modes), ZFP offers a fixed-rate mode \cite{zfp}.
The fixed-rate mode 
of ZFP offers  precise control over the number of bits per symbol in the input data.
It operates by transforming the data into a highly compressible domain and truncating each symbol to reach the appropriate rate.
However, the fixed rate mode of ZFP is not error bounded, and it suffers from significantly lower compression quality than does the fixed-accuracy mode of ZFP. Figure \ref{fig:hurricane-tcf-zfp-vis} (b) demonstrates the compression quality of ZFP using fixed-accuracy mode and fixed-rate mode, respectively. One can clearly see that the latter exhibits much worse rate distortion than does the former (up to 30 dB difference with the same bit-rate in most  cases). Rate distortion is an important indicator to assess the compression quality. Its detailed definition can be found in Section \ref{sec:exp-results} (4). Figures \ref{fig:hurricane-tcf-zfp-vis} (c) and (d) clearly show that the fixed-rate mode results in much lower visual quality (e.g., more data loss) than the fixed-accuracy mode with the same compression ratio,  50:1 in this example. In absolute terms, the fixed-accuracy mode leads to much higher peak signal-to-noise ratios (PSNR) and lower auto-correlation of compression errors (ACF(error)), which means better compression quality than the fixed-rate mode.
In ZFP's website and user guide, the developer of ZFP also points out that the fixed rate mode is not recommended unless one needs ``to bound the compressed size or need random access to blocks'' \cite{zfp}.

In contrast, not only does our framework fix the compression ratio but it also achieves higher compression quality for different compressors (such as SZ, ZFP, and MGARD) based on their error-bounding mode. Additionally, it can provide random access to the same level as can ZFP's fixed rate mode when supported by the underlying compressor. Since our framework utilizes a control loop to bound the compression ratio, it may suffer a lower bandwidth than ZFP's fixed-rate mode to a certain extent. The tradeoff for this lower bandwidth is compressed data of far higher quality for the same compression ratio, which we demonstrate in detail in the evaluation section.

The literature also includes studies investigating the use of fixed-ratio compressors for images. One such work \cite{kimFixedRatioCompressionRGBW2016} is JPEG-LS, a fixed-ratio compressor for images. This work adopts a combination of a prediction system for data values and two runs of Golumb-Rice encoding to encode RGB values for images. The first run of the Golumb-Rice encoding is used to estimate the quantization level used in the  the second run of the encoder.  Golumb-Rice assumes integer inputs, whereas our work is applicable on all numeric inputs.

Some work also has been done on fixed-ratio compression  in  digital signal processing \cite{andrewsAdaptiveDataCompression1967}
In this domain, adaptive sampling techniques are used to maintain a budget for how many points to transmit.
When a point is determined to provide new information (using a predictor, interpolation scheme, or some other method), it is transmitted and the budget is expended as long as there is remaining budget.
If the budget is spent, then no points are transmitted until the budget is refilled.
Over time, the budget is increased to keep the rate constant.
In contrast, our work does not rely on a control loop to maintain the error budget, so can look at the data holistically to decide where to place the loss in our signal, allowing for more accurate reconstructions.

\section{Problem Formulation}
\label{sec:formulation}

In this section, we formulate the research problem, by clarifying the inputs, constraints, and the target of our research. 

Before describing the problem formulation, however, we  introduce some related notations as follows.
Given a specific field $f$ at time-step $t$ of an application, we denote the dataset by $D_{f,t}$ = $\{d_1, d_2, \dots, d_n\}$, where $d_i$ refers to the original value of data point $i$ in the dataset and $n$ is the number of elements. 
We denote its corresponding decompressed dataset by $D'_{f,t}$= $\{d'_1, d'_2, \cdots, d'_n\}$, where $d'_i$ refers to the reconstructed value after the decompression.
We denote the original data size and compressed data size by $s(D_{f,t})$ and $s(D'_{f,t})$, respectively. The compression ratio (denoted by $\rho$) then can be written as $\rho(D_{f,t})$ = $\frac{s(D_{f,t})}{s(D'_{f,t})}$.
Moreover, we denote the target compression ratio specified by the users as $\rho_t(D_{f,t})$, and the real compression ratio after the compression as $\rho_r(D_{f,t}, e)$. 

The fixed-ratio lossy compression problem is formulated as follows, based on whether it is subject to an error-control constraint or not.

\begin{itemize}
    \item \textit{Nonconstrained fixed-ratio compression}: The objective of the nonconstrained fixed-ratio lossy compression is to confine the real compression ratio to be around the target compression ratio within a user-specified tolerable error (denoted by $\epsilon$), as shown below. 
\vspace{-2mm}
\begin{equation}
\label{ieq:target}
  \rho_t(D_{f,t})-\epsilon \leq\rho_r(D_{f,t})\leq \rho_t(D_{f,t})+\epsilon
  \vspace{-2mm}
\end{equation}    
    \item \textit{Error-control-based fixed-ratio compression}: The objective of the error-control-based fixed-ratio compression is to tune the compression ratio to be within the acceptable range [$\rho_t(D)$$-$$\epsilon$,$\rho_t(D)$$+$$\epsilon$], while respecting the user-specified error bound (denoted by $e$), as shown below.
    \vspace{-2mm}
\begin{equation}
  \label{ieq:constraint-target}
  \begin{array}{l}
    \rho_t(D_{f,t})-\epsilon \leq\rho_r(D_{f,t}, e)\leq \rho_t(D_{f,t})+\epsilon \\
    \hspace{5mm} s.t. \hspace{1mm} \Pi(D_{f,t},D'_{f,t})\leq e ,
  \end{array}
  \vspace{-2mm}
\end{equation}
    where $\Pi(D_{f,t},D'_{f,t})$ is a function of error control. For instance, $\Pi(D_{f,t},D'_{f,t})$=$\max_{i} |d_i-d'_i|$ for the absolute error bound, and $\Pi(D_{f,t},D'_{f,t})$=$\sum_{d_i\in D_{f,t},d'_i\in D'_{f,t}}{(d_i-d'_i)^2}$ for the mean squared error bound.
\end{itemize}

We summarize  the key notation in Table \ref{tab:notation}.

\begin{table}[ht]
\centering
  \caption{Table of Key Notation}
  \vspace{-2mm}
  \label{tab:notation}
\begin{adjustbox}{max width=0.48\textwidth}  
  \begin{tabular}{ | c | l | }
      \hline
      \textbf{Notation} & \textbf{Description} \\
      \hline
      $D$ & original data set for all time-steps and fields\\
      \hline
      $D_{f}$ & original data set for all time-steps of a particular field\\
      \hline
      $D_{f,t}$ & original data set for a particular field and time-step\\
      \hline 
      $D'_{f,t}$ & decompressed data set for a particular field and time-step \\ 
      \hline 
      $\rho_t$ & target compression ratio \\
      \hline 
      $\rho_r$ & real compression ratio\\
      \hline
      $\epsilon$ & acceptable error for $\rho_t$ \\
      \hline 
      $e$ & error bound for compression \\
      \hline
      $\gamma$ & maximum value of the loss function \\
      \hline
      $\theta$ & fixed parameters of the compressor \\
      \hline
      $N$ & the number of dimensions \\
      \hline
      $n$ & the total number of data points \\
      \hline
      $T$ & the number of time-steps \\
      \hline
      $\alpha$ & the degree overlap between error-bound search ranges  \\
      \hline
      $U$ & maximum allowed compression error\\
      \hline
 \end{tabular}
\end{adjustbox}
\end{table}

\section{Design and Optimization}
\label{sec:design-opt}

In this section, we present the design of our fixed-ratio lossy compression framework and optimization strategies.

\subsection{Design Overview}

\begin{figure} [ht]
\centering
  \includegraphics[scale=0.5]{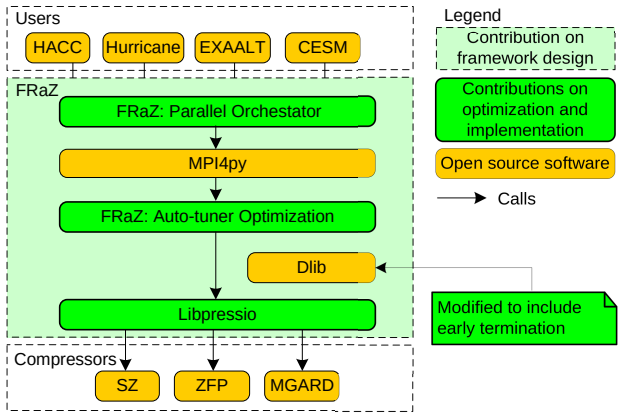}
  \caption{Design overview and summary of our contributions}
  \label{fig:contributions}
\end{figure}

Figure \ref{fig:contributions} shows the design overview with highlighted boxes indicating our major contributions in this paper and the relationship among different modules in the framework. As shown in the figure, our FraZ framework is composed of five modules, and the optimizing autotuner and parallel orchestrator are the core modules. They are, respectively, in charge of (1) searching for the optimal error setting based on the target compression ratio with few iterations and (2) parallelizing the overall tuning job involving different searching spaces for each field and different time-steps and across various fields. We develop an easy-to-use library (called Libpressio \cite{CODARcodeLibpressio2019}) to build a middle layer for abstracting the discrepancies of the APIs of different compressors.

We list our major contributions as follows:

\begin{enumerate}
    \item Formulated fixed-ratio compression as an optimization problem in a way that  converges quickly without resorting to multiobjective optimization
    \item Evaluated several different optimization algorithms to find one that works on all of our test cases, and then modified it to improve performance for our FRaZ
    \item Implemented and ran parallel search to improve the throughput of the technique
\end{enumerate}

\subsection{Autotuning Optimization}

In this subsection, we describe our autotuning solution in detail, which includes three critical parts: (1) exploration of the initial optimization  methods, (2) construction of a loss function, and (3) improvements to the optimization algorithm that involves how to deal with infeasible target compression ratio requirement and determines the exact error-bound setting.

\subsubsection{Exploration of Initial Optimization Methods}

In this subsection we describe how we choose which optimization method to use as a starting point for later refinement.

\begin{figure}[ht]
    \centering
    \includegraphics[scale=0.6]{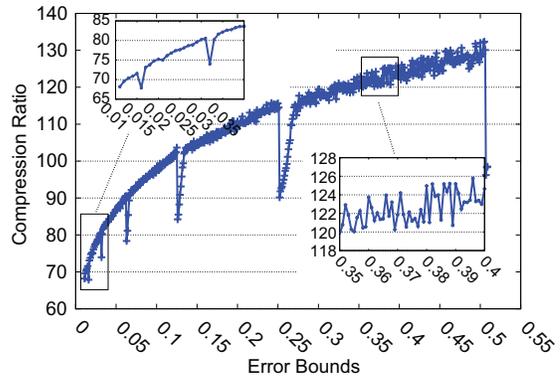}
    \vspace{-3mm}
    \caption{Example based on the hurricane simulation dataset (field: QCLOUDf.log10) showing that the relationship between error bounds and compression ratios is not always monotonic}
    \label{fig:hurricane_nonmonotonic}
\end{figure}

Before detailing FRaZ's optimizing autotuning method, we first analyze why the straightforward binary search is not suitable for our case.
On the one hand, the application datasets may exhibit a non-monotonic compression ratio increase with error bounds.
We present a typical example in Figure~\ref{fig:hurricane_nonmonotonic}, which uses SZ to compress the QCLOUDf field of the hurricane simulation dataset. We can clearly see that the compression ratios may decrease significantly with larger error bounds in some cases. We also observe the spiky changes in the compression ratios with increasing error bounds on other datasets (not presented here due to space limit).
The reason is that SZ needs to use decompressed data to do the prediction during the compression, which may cause unstable prediction accuracy. Moreover, SZ's fourth stage (dictionary encoder) may find various repeated occurrences of bytes based on output of the third stage, because a tiny change to the error bound may largely affect the Huffman tree constructed in the third phase of SZ. By comparison, our autotuning search algorithm is a general-purpose optimizer and takes into account the irregular relationship between compression ratios and error bounds. 
On the other hand, even on the datasets where monotonicity holds, binary search may still be slower than FRaZ's optimizing autotuner.
For example, when searching for the target compression ratio 8:1 at the $48^{th}$ time-step on the Hurriane-CLOUD field,
our method requires only 6 iterations to converge to an acceptable solution, whereas binary search needs 39 iterations.
The reason is that binary search may spend substantial time searching small error bounds, which would not result in an acceptable solution because it climbs from the minimum possible error bound to the user-specified upper limit.

When developing FRaZ's optimizing autotuner, we considered a number of different techniques to perform the tuning.
Since we are developing a generic method, we cannot construct a general derivative function that relates the change in error bound to the change in compression ratio.
Therefore, we need to decide between methods that use numerical derivatives and derivative-free optimization because the derivative of the compression ratio with respect to the error bound is unknown.
The methods using numerical derivatives approximate the slope of the objective function by sampling nearby points.
Some methods that fall in this category are gradient descent (i.e., Newton-like methods such as \cite{fischer1992special} and ADAM \cite{kingma2014adam}).
However, when evaluating an error bound to determine the compression ratio, we must run the compressor since we are using the compressors as black boxes, which may take a substantial amount of compared with the optimization problem.
In this sense, numerical derivative-based methods are too slow.

\begin{figure}
  \centering
  \includegraphics[scale=0.4]{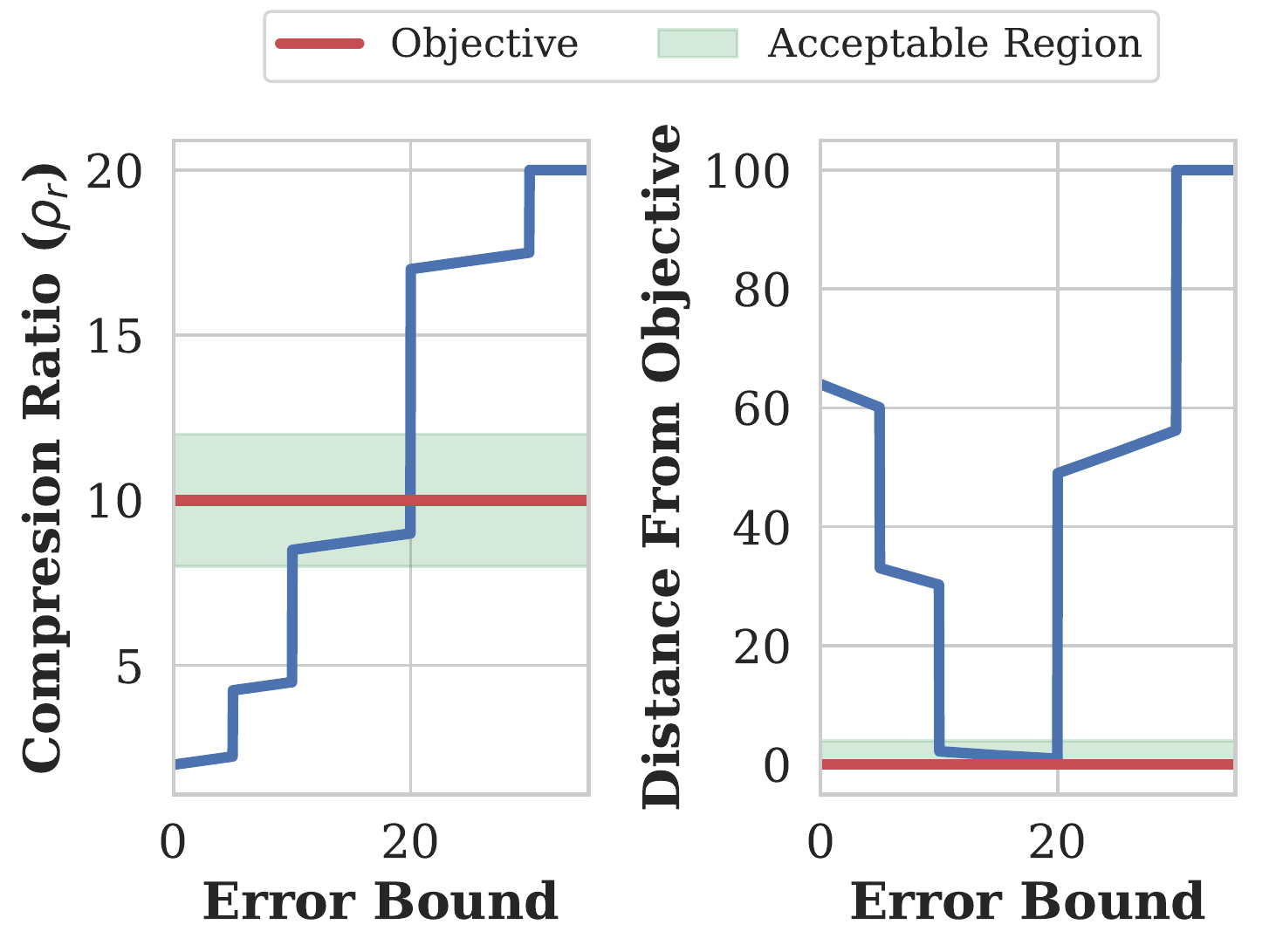}
  \vspace{-3mm}
  \caption{Illustration of autotuning optimization function:
  On the left is a hypothetical relationship between an error-bound level and compression ratio for some compressor and dataset.
  The target compression ratio is marked as a red line, and the acceptable region is colored green.
  On the right is the corresponding loss function using our method.
  The green area above the target compression ratio refers to the acceptable region.
  In this case, where there are blue points in the acceptable region, we call the result feasible.
  If the acceptable region was below the blue points, we would call it infeasible.
  }
  \label{fig:opt_overview}
\end{figure}

We therefore turned our consideration to derivative-free optimization.
We considered methods such as BOBYQA \cite{powell2009bobyqa}, but they do not handle a large number of local optimums. This ability  is essential for developing a robust tuning framework for lossy compression because many of the functions that relate error bounds to compression ratios look like the plot on the left of Figure~\ref{fig:opt_overview}: a step-like function with perhaps a slight upward slope on each step.
In practice, we noted that it is easily able to escape the local optima in these functions.

We also took into account a variety of implementations of these algorithms, such as the ones in \cite{saltzman2002coin, gabriele1977optlib, kingDlibLibraryOptimization2018}.
We decided between these libraries using three criteria: (1) correctness of the result, (2) time to solution, and (3) modifiability and readability of code.
Ultimately we started with a black-box optimization function called \texttt{find\_global\_min} from the commonly used Dlib library from which we make our modifications \cite{kingDlibLibraryOptimization2018}.
The global-minimum-finding algorithm designed by Davis King that combines the works of \cite{malherbeGlobalOptimizationLipschitz2017} and \cite{powellNEWUOASoftwareUnconstrained2006}.
It requires a deterministic function that maps from a vector to a scalar, a vector of lower bounds, and a vector of upper bounds as inputs.
At a high level the algorithm works as follows.
It begins with a randomly chosen point between the upper and lower bounds.
Then, it alternates between a point chosen by the model in \cite{malherbeGlobalOptimizationLipschitz2017}, which  approximates the function by using a series of piecewise linear functions and chooses the global minimum of this function,
and the model in \cite{powellNEWUOASoftwareUnconstrained2006}, which  does a quadratic refinement of the lowest valley in the model.
According to \cite{kingDlibLibraryOptimization2018}, this method performs well on functions with a large number of local optimums, and this performance was confirmed by our experience.

\subsubsection{Construction of Loss Function}

Now that we have an optimizer framework, we need to construct a loss function.
First, we created a closure for each compressor, $\rho_r(D_{f,t}, e)$ that transformed its interface including a dataset $D$ and parameters $\theta$ in a function accepting only the error bound $e$.
To create the closure, we developed libpressio \cite{CODARcodeLibpressio2019}---a generic interface for lossy compressors that abstracts between their differences so that we could write one implementation of the framework for SZ, ZFP, and MGRAD.

To convert this to a loss function, we chose the distance between the measured compression ratio and target compression ratio $\rho_r(D_{f,t}, e) - \rho_t(D_{f,t})$.
Now, the function that relates an error bound to a compression ratio is an arbitrary function that may or may not have a global or local optimum.
Therefore, we transformed the function by applying a clamped square function (i.e. $\min \left( x^2, \gamma \right)$, where $\gamma$ is equal to 80\% of the maximum representable double using IEEE 754 floating-point notation).
This maps the possible range of the input function from the range $\left( -\infty, \infty \right)$ to the range $\left[ 0 , \gamma \right]$.
The benefits of this are twofold.
First, the function now has a lowest possible global minimum we can optimize for.
Second, the function now has a highest possible value that avoids a bug in the Dlib \texttt{find\_global\_min} function that causes a segmentation fault.
We also considered the function $\min \left( |x|, \gamma \right)$, but found that the quadratic version converged faster.
This leaves us with the final optimization function $l(e) = \min \left( \left(\rho_r(D_{f,t},e) - \rho_t(D_{f,t}))\right)^2, \gamma \right)$.

\subsubsection{Development of Worker Task Algorithm}

Our next insight was that often the exact match of the compression ratio is not always feasible and is neither desired nor required.
It may not always be feasible because for some compressors, for example ZFP's accuracy mode, the function that maps from the error bound to the compression ratio is a step function, such that not all compression ratios are feasible. In addition, it may not always be desired or required because the user might accept a range of compression ratios and prefer finding a match quickly rather than waiting for a more precise match.

Looking again at Figure~\ref{fig:opt_overview}, we see a typical relationship between an error bound and the compression ratio.
If the user asks for a compression ratio of 15, no error bound  would satisfy that request using this compressor.
In contrast, FRaZ will return the closest point that it observes to the target; in the case of Figure~\ref{fig:opt_overview} it would report an error bound that results in a compression ratio near 17.5.
Depending on the user's global error tolerance, this value near 17.5 may or may not be within the user's acceptable region, meaning it may or may not be a feasible solution.

Another case that the solution may be infeasible is when needed error bound required to meet the objective is above the user's specified upper error bound, $U$.
In this case, FRaZ will report the error bound that resulted in the closest that it observed to the target compression ratio, and the user can run FRaZ again with the default upper bound, which is equal to the maximum allowed level of an error bound by the compressor.
If FRaZ identifies a solution in this case, the user can evaluate whether to relax the perhaps overly strict error tolerance to meet the objective or decide that the fidelity of the results is more important and that the bound cannot be relaxed.
Alternatively, the user can try a different compressor backend that implements the same error bound.

In fact, determining the exact error bound that produces a specified compression ratio may not be desired or required.
The reason is that a large number of iterations may be needed in order to converge to an error bound, and the user would rather trade time for accuracy.
Therefore, we implemented a version of Dlib's \texttt{find\_global\_min} that implements a global cutoff parameter $\epsilon \in [0,1]$. 
Specifically, we allow the algorithm to terminate if the result of the optimization function results in a value in the range:
$[0,\epsilon^2 \rho_t(D_{f,t})^2]$.
This has a substantial impact on the performance on the typical case.

We combine these insights into our worker task algorithm, as shown in Algorithm~\ref{alg:worker_task}.

\begin{algorithm}
\footnotesize
\caption{\textsc{Worker Task}} \label{alg:worker_task}
\renewcommand{\algorithmiccomment}[1]{/*#1*/}
  \textbf{Input}: target ratio ($\rho_t(D_{f,t})$), acceptable error $\epsilon$, dataset $D_{f,t}$, prediction $p$, region's lower bound $l$, region's upper bound $u$ \\
  \textbf{Output}: real compression ratio $\rho_r(D_{f,t}, e)$, recommended error bound setting $e$

\begin{algorithmic}[1]
  \IF{$p \ne 0$}
    \STATE $\rho_r(D_{f,t}, e) \gets compress(D_{f,t}, p)$ \COMMENT{If a prediction was provided, try it first.}
  \ENDIF
  \IF{ $(1-\epsilon) * \rho_t(D_{f,t}) \le \rho_r(D_{f,t}, e)) \le (1+\epsilon) * \rho_t(D_{f,t})$}
    \RETURN $\rho_r(D_{f,t}), p$ \COMMENT{terminate if $\rho_r(D_{f,t}, e)$ meets requirement.}
  \ENDIF
  \STATE $\rho_r(D_{f,t}, e), e \gets train\_with\_cutof\hspace{-0.5mm}f(D_{f,t}, l, u, \rho_t(D_{f,t}), \epsilon)$
  \RETURN $\rho_r(D_{f,t}, e), e$
\end{algorithmic}
\end{algorithm}

\subsection{Parallelism Scheme}

After optimizing the serial performance via the design of the optimization algorithm, we develop a parallel optimization method using Dlib's built-in multithreaded optimization mode.
Some compressors (such as SZ and MGARD) do not support being run with different settings in a multithreaded context because of the use of global variables.
In this situation, we can only treat each compression as a non-multithreaded task because we are developing a generic framework.

\begin{figure}
    \centering
  \includegraphics[width=0.9\columnwidth]{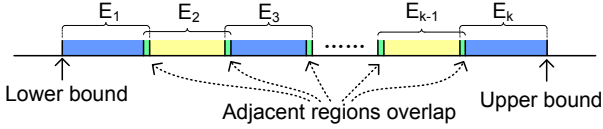}
  \vspace{-2mm}
  \caption{Illustration of error bound ranges.  We divide the range from lower bound to upper bound into $K$ slightly overlapping regions ($E_1, E_2, \dots E_K$).
  The overlap is a small fixed percentage of the width of the regions (i.e., 10\%).
  Each region ($E_1$, $E_2$, $\dots$, $E_K$) is then passed from the parallel orchestrator to the autotuning optimizer.
  Note that the ends $E_1$ and $E_K$ are slightly smaller to preserve the error bound range.
  }
  \label{fig:errorboundrange}
\end{figure}

We use multiple processes based on MPI to parallelize the search by error bound range.
Figure~\ref{fig:errorboundrange} provides an overview of our method.
Rather than a serial search over the entire lower to upper bound range, we  divide the range into $k$ overlapping regions.
We then give each of the $k$ regions to separate MPI processes, and use Algorithm~\ref{alg:parallel_eb_task} to process them.
As the processes complete, we test whether we have satisfied our objective subject to our global threshold $\epsilon$ (line 7--9).
If so, we terminate all tasks that have not yet begun to execute (line 10--14).
If a particular task finishes and we have not satisfied our objective, we do nothing.
If all the tasks finish and we still have not met our objective, we conclude that the requested compression ratio is infeasible (line 18--25).

So why do we overlap the error bound regions?
Overlapping the regions avoids extremely-long worst-case search time in the optimization algorithm.
Since we terminate early once a solution is found, FRaZ’s runtime depends on the region containing the target.
Without small overlapping, if the target error-bound coincides a border, its MPI\_rank iterates longer lacking stationary-points for quadratic refinement.

\begin{algorithm}
\footnotesize
\caption{\textsc{Training}} \label{alg:parallel_eb_task}
\renewcommand{\algorithmiccomment}[1]{/*#1*/}
  \textbf{Input}: target compression ratio $\rho_t(D_{f,t})$, acceptable error $\epsilon$, dataset $D_t$, max allowed compression error $U$ \\
\textbf{Output}: real compression ratio $\rho_r(D_{f,t}, e)$, recommended error bound setting $e$

\begin{algorithmic}[1]
  \STATE $tasks[N]$
  \STATE $done \gets false$
  \FOR{$(i, (l,u)) \in make\_error\_bounds(U)$}
    \STATE $tasks[i] \gets launch\_task(D_t, l, u, \rho_t(D_{f,t}), \epsilon, h)$
  \ENDFOR
  \WHILE{$not done$}
    \STATE $last\_task \gets next\_completed(tasks)$
    \STATE $candiate \gets compression\_ratio(last\_task)$
    \IF{$ \rho_t(D_{f,t}) (1-\epsilon) \le candidate \le \rho_t(D_{f,t}) (1+\epsilon) $}
      \STATE $done \gets true$
      \FOR{$task \in tasks$}
        \STATE $cancel\_if\_not\_finished(task)$
      \ENDFOR
    \ENDIF
    \STATE $done \gets has\_next(completed)$
  \ENDWHILE
  \STATE $\rho_r(D_{f,t}, e) = \infty$
  \FOR{$task \in tasks$}
    \IF{finished(task)}
      \STATE $\rho \gets compression\_ratio(task)$
      \IF{ $(\rho_r - \rho)^2 < (\rho_t - \rho)^2$}
        \STATE $\rho_r=  \rho$
      \ENDIF
    \ENDIF
  \ENDFOR
  \RETURN $\rho_r(D_{f,t}, e), error\_bound(task)$
\end{algorithmic}
\end{algorithm}

To limit the effects of waiting on wrong guesses, we constrain the number of iterations to a maximum value.
We considered limiting by time instead, but we were unable to find a heuristic that worked well across multiple datasets, fields, and time-steps.
This is because the compression time is a function of the dataset size, the entropy of the data contained within, and properties of each compressor.

Limiting the amount of wasted computational resources is desirable.
Since we are dividing on error-bound range, a small number of the searches (typically one) are expected to return successfully if the requested ratio is feasible.
Additionally, there seems to be a floor for how many iterations are required to converge for a particular mode of a compressors.
Hence, there is limited benefit to splitting into more than a few ranges, and cores could perhaps be more efficiently used for other fields.
Preliminary experiments found that 12 tasks per a particular field and time-step dataset offered an ideal tradeoff between efficiency and runtime, and we set it as the default.
The user can choose to use more tasks, however.

One can also perform additional optimization of multiple time-step data.
Often, subsequent iterations in a large simulation do not differ substantially and have similar compression properties.
Therefore we ran the first time-step as before, but then we assumed that the error bound found by the previous iteration was correct for the next full dataset.
If our assumption proved correct, we continued on and skipped training. Otherwise, we reran the training and adopted the new trained solution for the next step.
We then repeated this process over the remaining datasets.
In practice, we  retrained only a small percentage of the time.
On the hurricane dataset, for example, we retrained only 4 times on the CLOUD field.

We also take advantage of the embarrassingly parallel nature of parallelizing by fields, as shown in Algorithm~\ref{alg:parallel_field}. The results show some additional speedup.

\begin{algorithm}
\footnotesize
\caption{\textsc{Parallel by Field}} \label{alg:parallel_field}
\renewcommand{\algorithmiccomment}[1]{/*#1*/}
\textbf{Input}: target ratio $\rho_t(D_{f,t})$, $\epsilon$, dataset $D$, max allowed compression error $U$ \\
\textbf{Output}: real compression ratio $\rho_r(D_{f,t}, e)$, recommended error bound setting $e$

\begin{algorithmic}[1]
  \FOR{$D_f \in D$ \algorithmiccomment{in parallel}}
    \STATE $p \gets 0 $
    \FOR{ $D_{f,t} \in D_f$ }
      \STATE $\rho_r(D_{f,t}, e), e \gets parallel\_error\_bound(D_t, \epsilon, U)$
      \IF{$(1-\epsilon) * \rho_t(D_{f,t}) \le \rho_r(D_{f,t}, e) \le (1+\epsilon) * \rho_t(D_{f,t})$}
        \STATE $p \gets e$
      \ENDIF
    \ENDFOR
  \ENDFOR
\end{algorithmic}
\end{algorithm}

\section{Performance and Quality Evaluation}
\label{sec:evaluation}

In this section, we first describe our experimental setup, including hardware, software, and datasets. We then describe our evaluation metrics and results using five real-world scientific floating-point datasets on Argonne's Bebop supercomputer \cite{bebop}.

%
%
%

\subsection{Experimental Setup}
\subsubsection{Hardware and Software Used for Evaluation}

The hardware and software versions we used on the Bebop supercomputer \cite{bebop} are given in Table~\ref{tab:hardwaresoftware}.

    \vspace{-2mm}
\begin{table}[ht]
    \centering
    \caption{Hardware and Software Versions Used}
    \vspace{-2mm}
    \begin{adjustbox}{max width=\columnwidth}     
    \begin{tabular}{|ll|ll|}
        \hline
        \textbf{Hardware} & \multicolumn{3}{l|}{\textbf{Description}} \\
        \hline
        CPU & \multicolumn{3}{l|}{36 Core Intel Xeon E5-2695v4} \\
        \hline
        MEM & \multicolumn{3}{l|}{128GB DDR4 Ram}\\
        \hline
        NIC & \multicolumn{3}{l|}{Intel Omni-Path HFI Silicon 100 Series}\\
        \hline
        \hline
        \textbf{Software} & \textbf{Description} & \textbf{Software} & \textbf{Description}\\
        \hline
        OS & CentOS 7 & SZ & 2.1.7\\
        \hline
        CC/CXX & gcc/g++ 8.3.1& ZFP & 0.5.5\\
        \hline
        MPI & OpenMPI 2.1.1& MGARD & 0.0.0.2\\
        \hline
        Dlib & 2.28& Singularity & 3.0.2\\
        \hline
    \end{tabular}
    \end{adjustbox}
    \label{tab:hardwaresoftware}
\end{table}

We have packaged our software as a Singularity container for reproducibility.

\subsubsection{Datasets used for Experiments}

In our experiments, we evaluated our designed fixed-ratio lossy compression framework based on all three state-of-the-art compressors described in Section \ref{sec:background}, using five real-world scientific simulation datasets downloaded from scientific data reduction benchmark \cite{sdrb}. The raw data are all stored in the form of single-precision data type (32-bit floating point). We describe the five application datasets in Table~\ref{tab:app}. 

\vspace{-.2cm}
\begin{table}[ht]
    \centering
    \caption{Dataset Descriptions}
\vspace{-.2cm}
    \begin{adjustbox}{max width=\columnwidth}      
    \begin{tabular}{|llcccc|}
        \hline
        \textbf{Name} & \textbf{Domain} &  \textbf{ \# Time-steps} & \textbf{Dim.} & \textbf{\# Fields}  & \textbf{Total size}\\
        \hline
        Hurricane & Meteorology & 48 & 3 & 13 & 59 GB\\
        \hline
        HACC & Cosmology& 101 & 1 & 6 & 11 GB\\
        \hline
        CESM & Climate & 62 & 2 & 6$^*$  & 48 GB\\
        \hline
        Exaalt & Moledular Dyn. & 82 & 1 & 3 & 1.1 GB\\
        \hline
        NYX & Cosmology & 8 & 3 & 5 & 35 GB\\
        \hline
        \multicolumn{6}{l}{* A limited number of fields had multi-time step data.  Only fields for which}\\
        \multicolumn{6}{l}{\hspace{2mm} multiple time step data were included.} \\
    \end{tabular}
    \end{adjustbox}
    \label{tab:app}
    \vspace{-2mm}
\end{table}

We chose these datasets  for a few reasons:
First, they offer results over multiple time-steps, which matches well user's practical post-analysis with a certain simulation period.
Second, the datasets use floating-point data which are often not served well by traditional lossless compressors.
Third, the datasets are commensurate with the use cases of fixed-ratio compression described in Section~\ref{sec:background}.

In some cases, we are not able to use all the datasets with all compressors.
We run all the experiments for all datasets and compressors when possible.
MGARD supports only 2d and 3d data so it is not tested on the HACC and Exaalt datasets. We adopt 6 typical fields for CESM application because other fields exhibit similar results with one of them (CLDHGH CLDLOW, CLOUD, FLDSC, FREQSH, PHIS).
We generally noted similar results for each dataset and compressor. 

\subsection{Experimental Results}
\label{sec:exp-results}

Over the course of our experiments, we evaluated four properties of FRaZ using the datasets from SDRBench \cite{sdrb}:

\begin{enumerate}
    \item How close do we get to the target compression ratio when it is feasible?
    \item How long does it take to find the target compression ratio or determine that it is infeasible?
    \item How does the runtime of the algorithm scale as the number of cores increase?
    \item How does FraZ compare with existing fixed-rate methods in terms of rate distortion and visual quality?
\end{enumerate}

\subsubsection{How close do we get to the target compression ratio?}

\begin{figure}[ht]
  \includegraphics[width=\columnwidth]{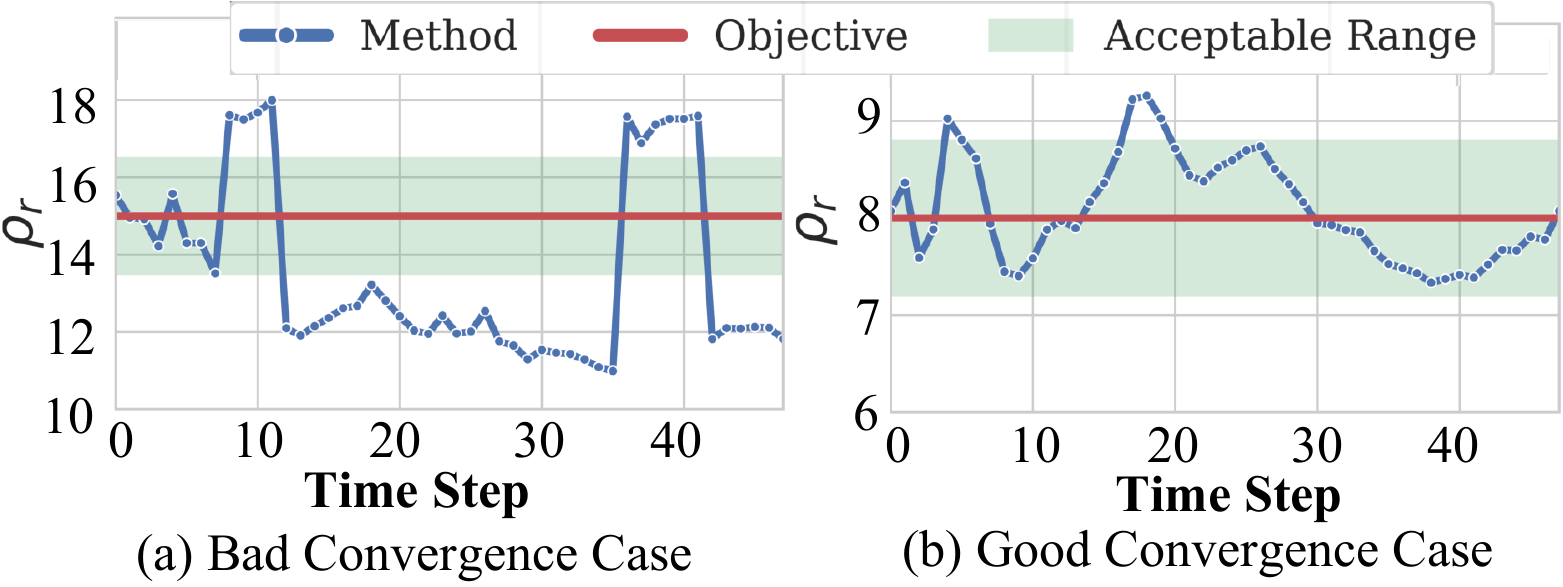}
  \vspace{-5mm}
  \caption{Demonstration of two types of convergence cases (Hurricane-CLOUD)}
  \label{fig:convergence-case}
\end{figure}

How close we get to the target compression ratio depends heavily on whether the requested compression ratio is feasible for the underlying compressor used.
Figure \ref{fig:convergence-case} (a) and Figure \ref{fig:convergence-case} (b) show a bad case and a good case, respectively.

In Figure~\ref{fig:convergence-case} (a), we see an example of where $\rho_t(D_{f,t})$ is infeasible for most time-steps for the CLOUD field.
The early time-steps compress within the acceptable range, but by time-step ten the $\rho_t(D_{f,t}) = 15$ is no longer feasible.
The reason  is that as the time-steps progress, the properties of the dataset change, affecting the ability of compressor to compresses it at this level.
As a result, we oscillate between a compression ratio that is larger and a compression ratio that is smaller.
However, a larger tolerance (i.e., $\epsilon = .2$) would have allowed even this case to converge for all time-steps.

In Figure~\ref{fig:convergence-case} (b), we see an example of where the algorithm converges on over 90\% of the time-steps.
In this case, we quickly converge to the acceptable range and are able to often reuse the previous time steps error bound for future iterations.
In this particular case, we have to retrain only four times over the course of the simulation on iterations: 0, 8, 15, 29.
Thus, the algorithm can quickly process many time-steps.

\subsubsection{How long does it take to reach the target compression ratio?}

\begin{figure}
  \includegraphics[width=\columnwidth]{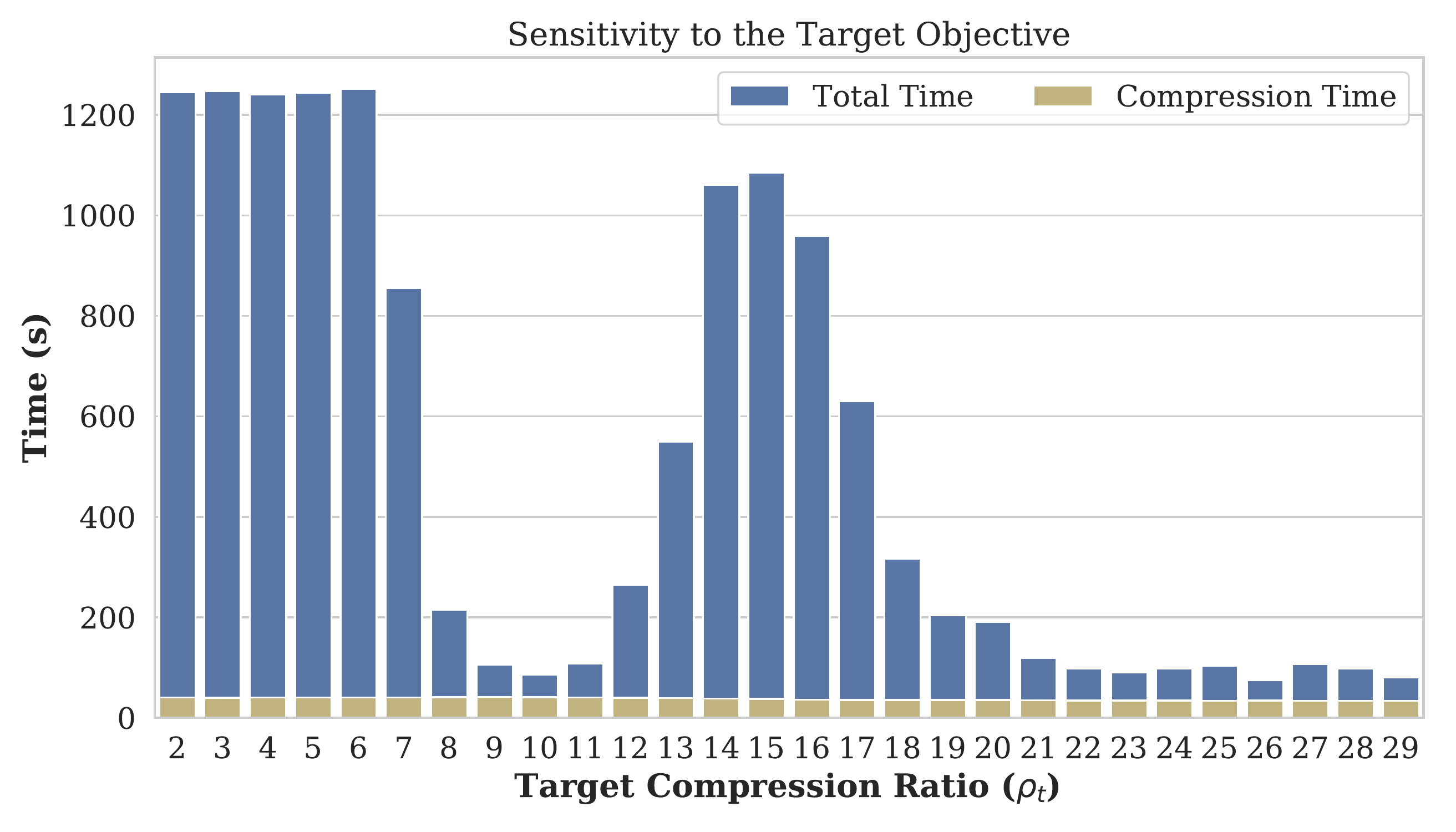}
  \vspace{-7mm}
  \caption{Sensitivity of FRaZ  to the choice of $\rho_t(D_{f,t})$:
    This is because not all values of $\rho_t(D_{f,t})$ are elements of the co-domain of the function that relates $\rho$ and $e$.
  }
  \label{fig:sensativity}
\end{figure}

When evaluating the algorithm, we wanted to consider how long the algorithm takes to find the target compression ratio.
This again depends greatly on whether $\rho_t(D_{f,t})$ is feasible or not.
Therefore, we considered a large number of possible  $\rho_t(D_{f,t})$'s for different datasets.
The results of this search are shown in Figure~\ref{fig:sensativity}. We can see that some compression ratios require far longer total times.
Figures~\ref{fig:convergence-case} (a) and (b) show a zoomed in view of $\rho_t(D_{f,t})=8$ and $\rho_t(D_{f,t})=15$.
The difference in runtime is explained by the difference in the number of time-steps that converge.
In the case shown in Figure \ref{fig:convergence-case} (a) relatively few time-steps converged because the objective was infeasible with the specified compressor; in the case shown in Figure ~\ref{fig:convergence-case} (b) almost all time-steps converged because the objective was feasible.
This resulted in about a 10x difference in performance between the two cases.

Why do low target compression ratios have long runtimes?
Many of the lossy compressors have an effective lower bound for the compression ratio.
In Figure~\ref{fig:sensativity}, it is about 7.5.
This effective lower bound on the compression ratio, means that FRaZ will never meet its objective and spends the remainder of the time searching until it hits its timeout.

How does this change across datasets?
In general, the more feasible compression ratios near the target, the better FRaZ preformed.
Each dataset had a compressor which was able to more accurately compress and decompress the data.

How does this change between compressors?
Generally SZ took less time than ZFP or MGARD even though ZFP may take less time for each compression.
This is because ZFP typically had fewer viable compression ratios than SZ due to limitations of ZFP's transform based approach.
As a result, FRaZ took more time-steps which took the maximum number of iterations lengthening the total runtime.
The difference in runtime between SZ and ZFP for a representative dataset can be seen in Figure~\ref{fig:scalability} below.


\subsubsection{How does the algorithm scale?}

To evaluate how well the algorithm scales, we considered the runtime of the algorithm as it scales over multiple cores on ANL Bebop \cite{bebop}.

\begin{figure}
    \centering
    \includegraphics[scale=0.45]{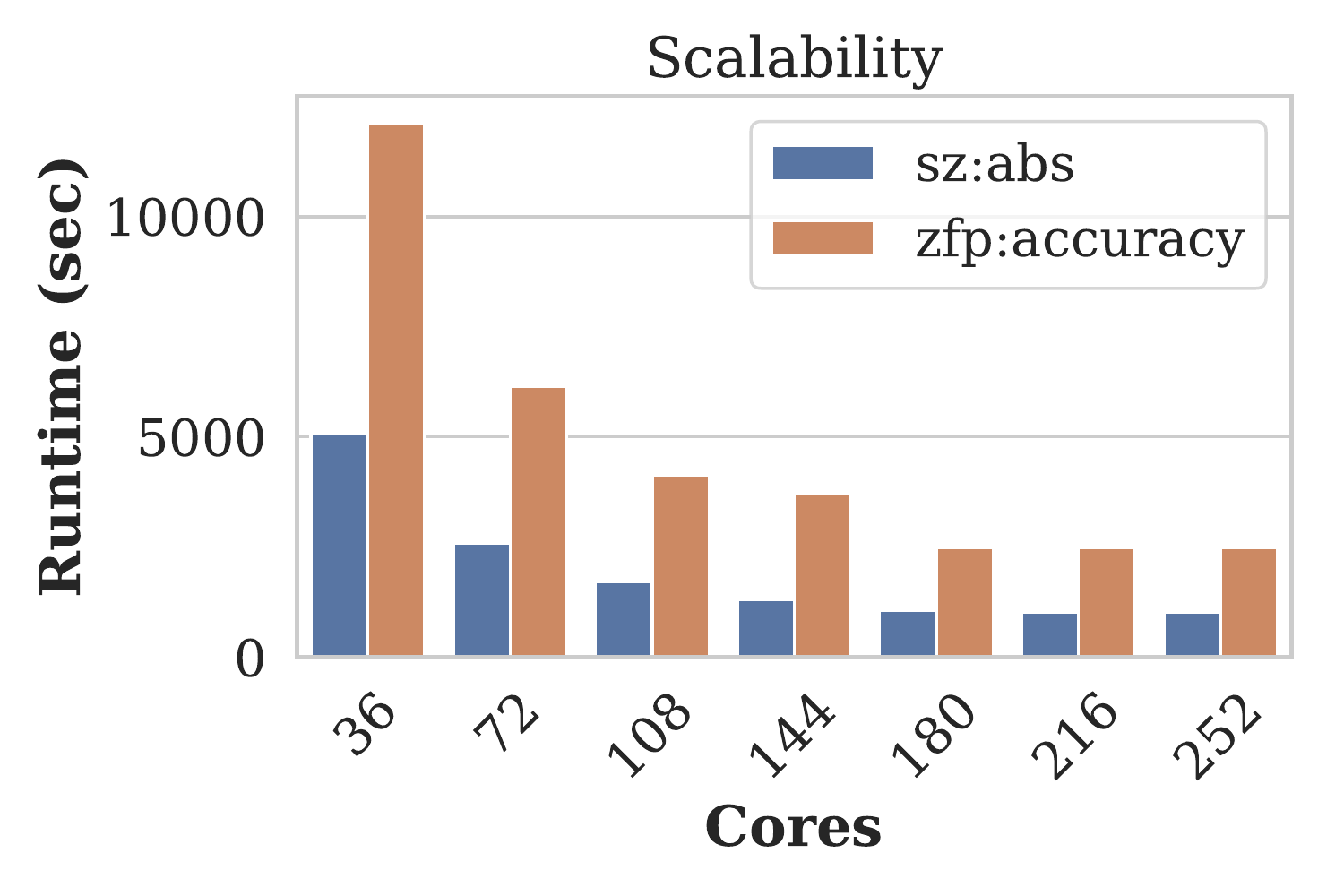}
    \vspace{-5mm}
    \caption{Scalability: our solution reaches the optimal performance at 180--216 cores, in that the total time is equal to the longest task's wallclock time at this scale.}
    \label{fig:scalability}
\end{figure}

Figure~\ref{fig:scalability} shows the strong scalability of the algorithm.
We see that  the algorithm scales 
by time-step and field levels for the first 180--216 cores with steep decreases in runtime due to parallelism at early levels and then less additional parallelism after that.
This is because the runtime of the algorithm is lower bounded by the longest running worker task.
All of the datasets we tested has at least one fields that takes substantially longer to compress than others.
And the scalability of this algorithm is limited by the longest of these.
In the case of the Hurricane dataset using the error-bounded compressor SZ, the QCLOUD field took 1022 seconds to compress while the 75 percentile is less than 500 and the 50 percentile is less than 325.

What accounts for the substantial difference between the scalability of FRaZ using ZFP and SZ?
Those familiar with ZFP likely know that it is typically faster than SZ, but this seems to contradict the result in Figure~\ref{fig:scalability}.
This result is explained by considering the individual fields rather than the overall scalability.
For the cases in which ZFP finds an error bound that satisfies the target compression ratio, it is much faster.
However ZFP often expresses fewer compression ratios for the same error bound range, resulting in more infeasible compression ratios and thus increasing the runtime.
ZFP expresses few compression ratios because it  uses a flooring function in the minimum exponent calculation used in fixed-accuracy mode.

\begin{figure} \centering
\hspace{-5mm}
\subfigure[{Hurricane(TCf48)}]
{
\raisebox{-1cm}{\includegraphics[scale=0.37]{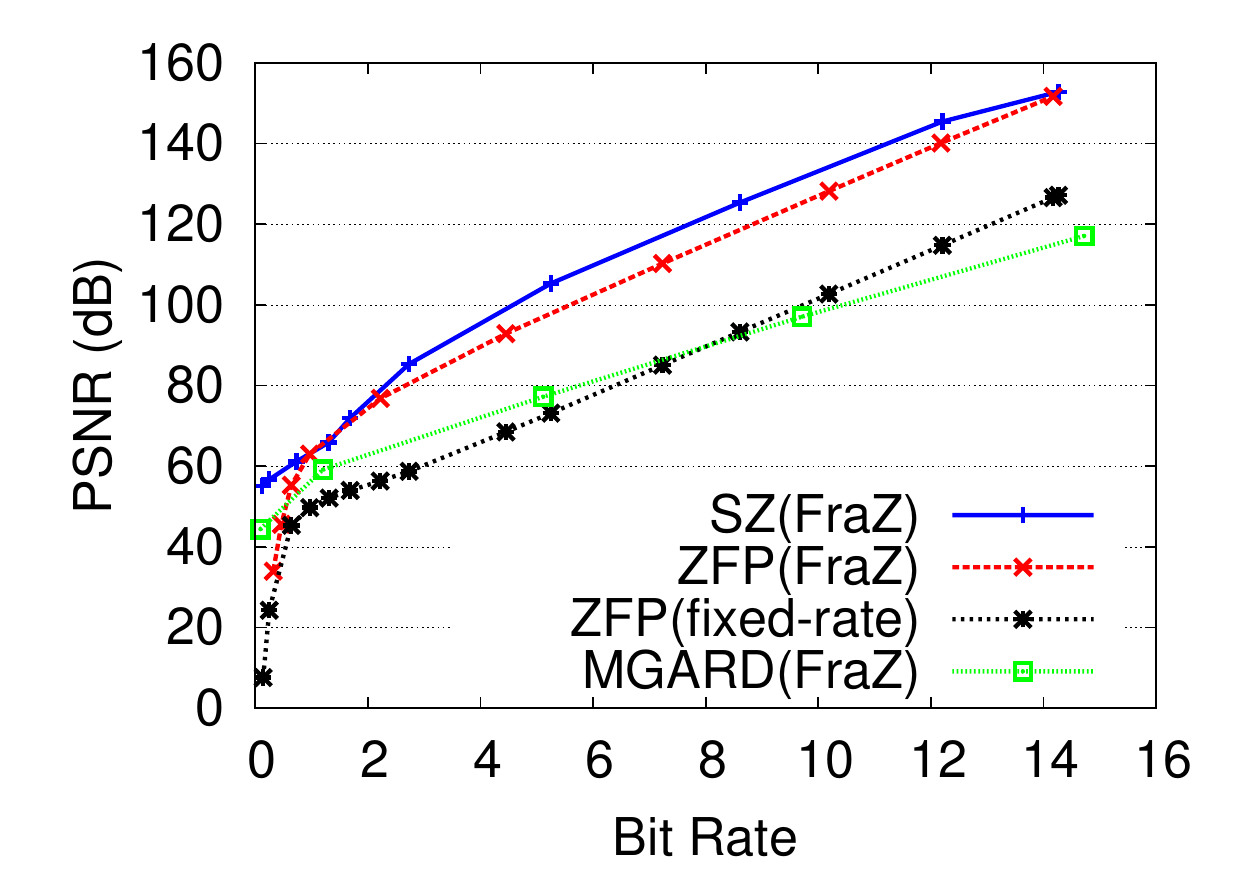}}
}
\hspace{-6mm}

\hspace{-5mm}
\subfigure[{NYX(temperature)}]
{
\raisebox{-1cm}{\includegraphics[scale=0.37]{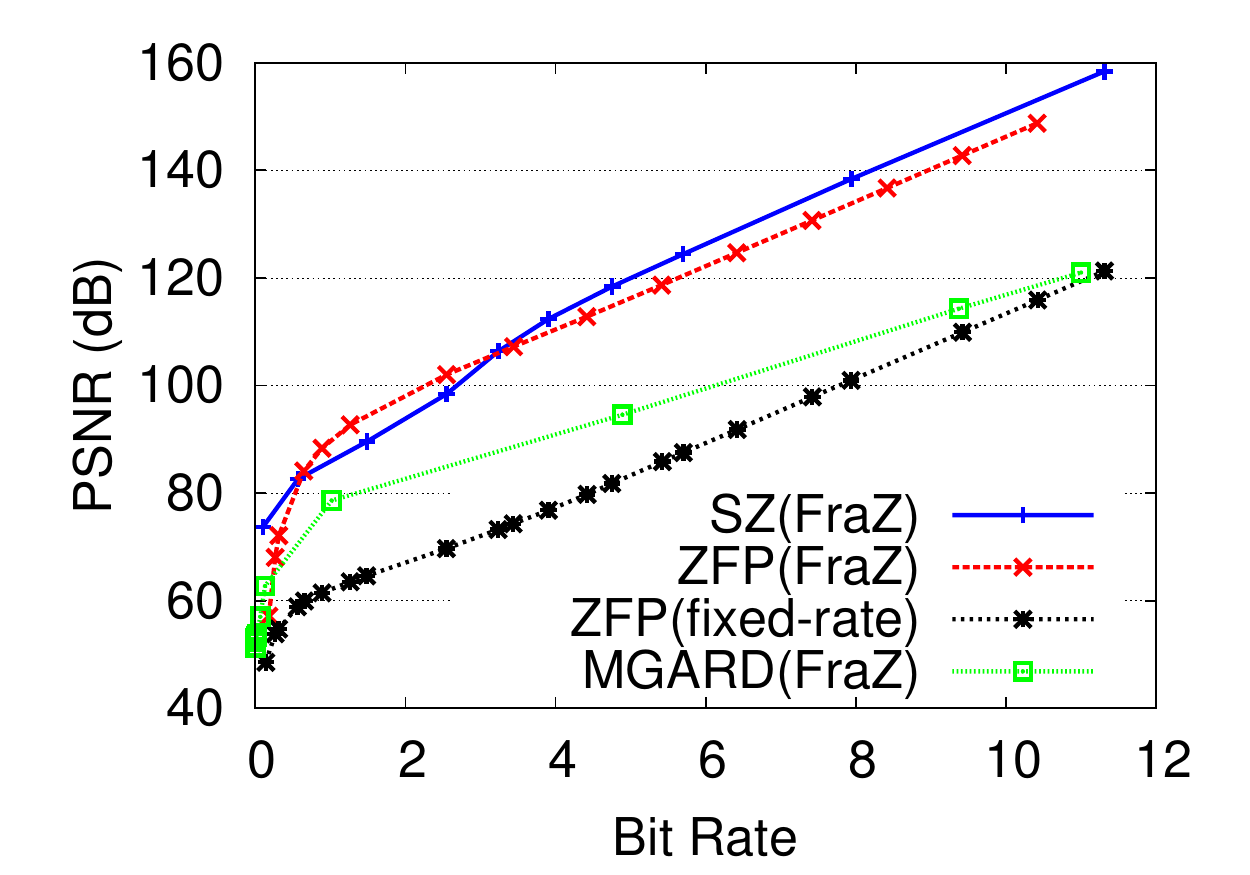}}
}
\hspace{-10mm}
\subfigure[{CESM-ATM(CLDHGH)}]
{
\raisebox{-1cm}{\includegraphics[scale=0.37]{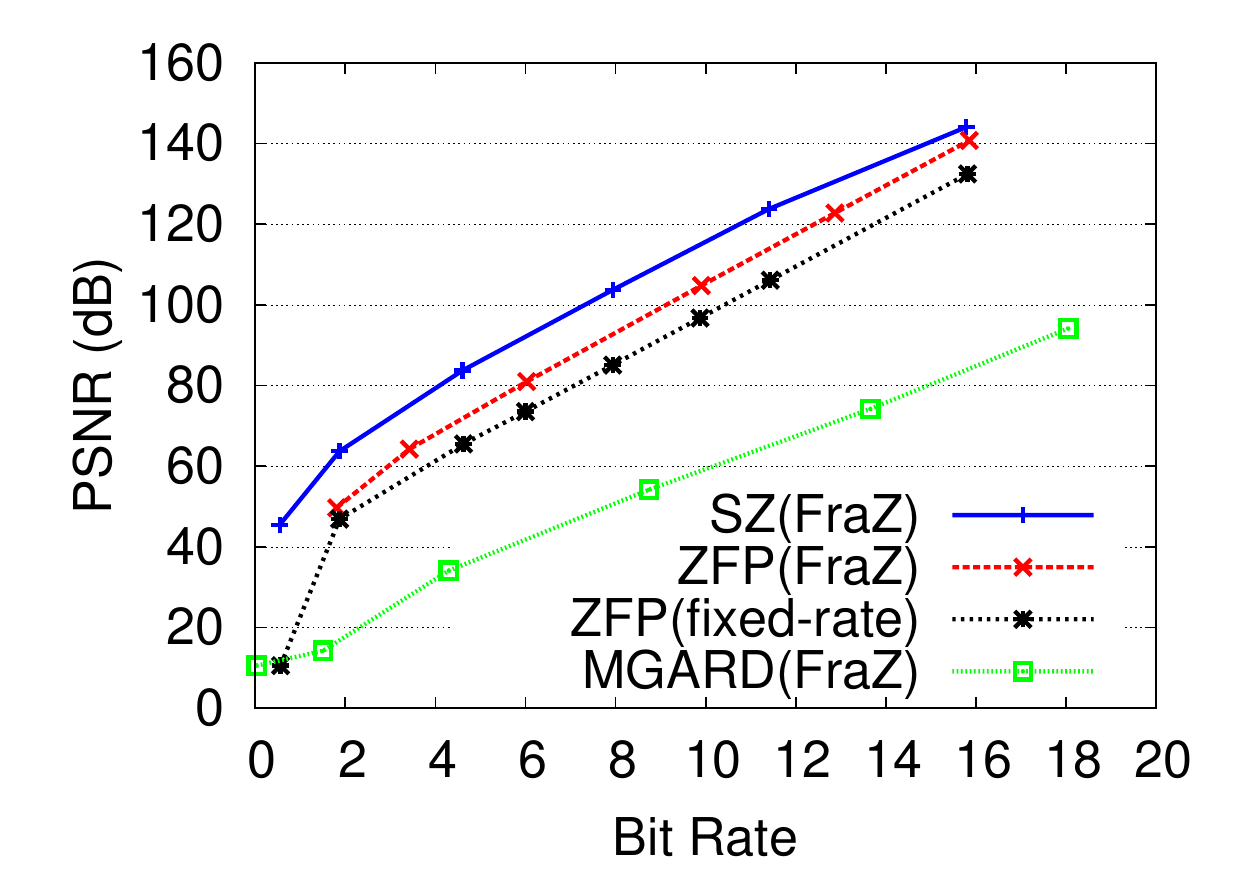}}
}
\hspace{-6mm}

\hspace{-5mm}
\subfigure[{HACC(x,y,z)}]
{
\raisebox{-1cm}{\includegraphics[scale=0.37]{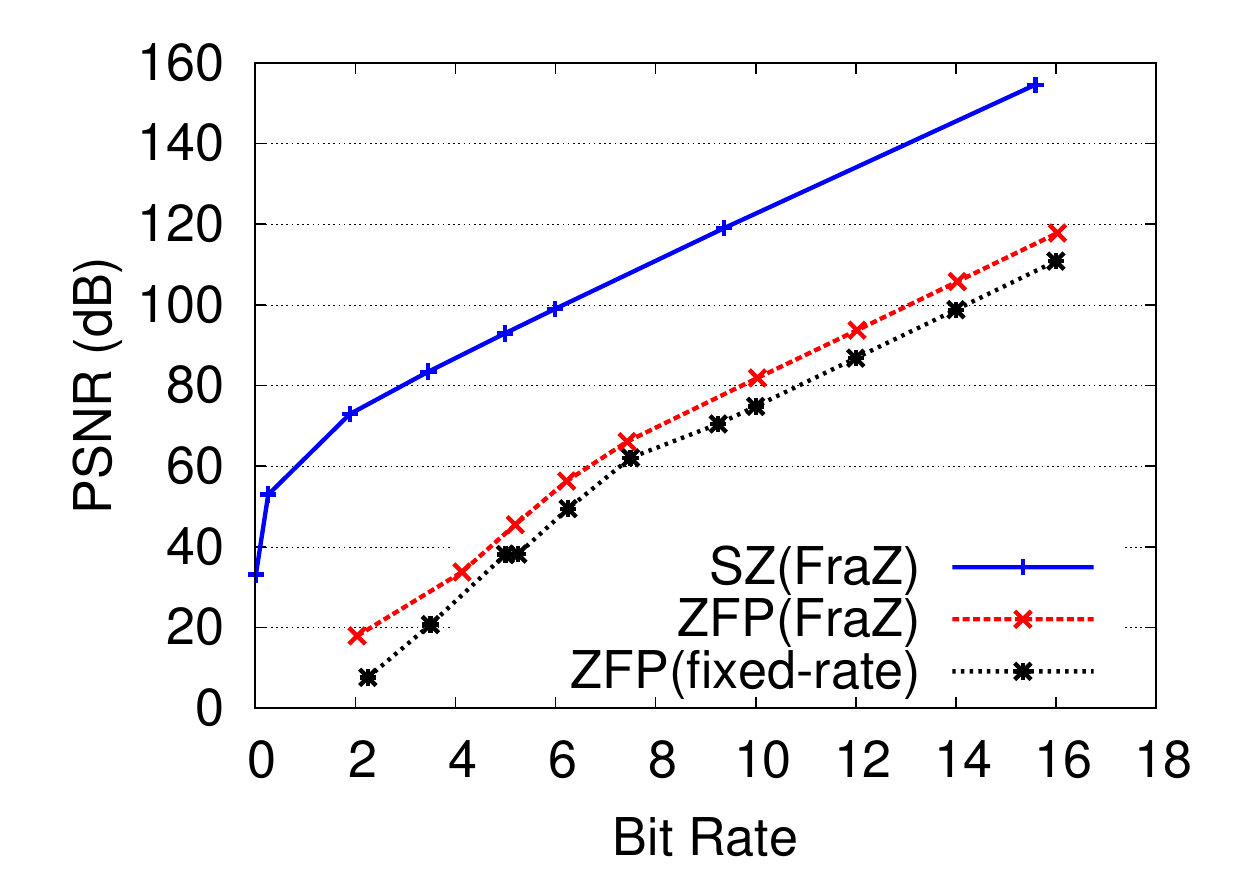}}
}
\hspace{-10mm}
\subfigure[{EXAALT(x,y,z)}]
{
\raisebox{-1cm}{\includegraphics[scale=0.37]{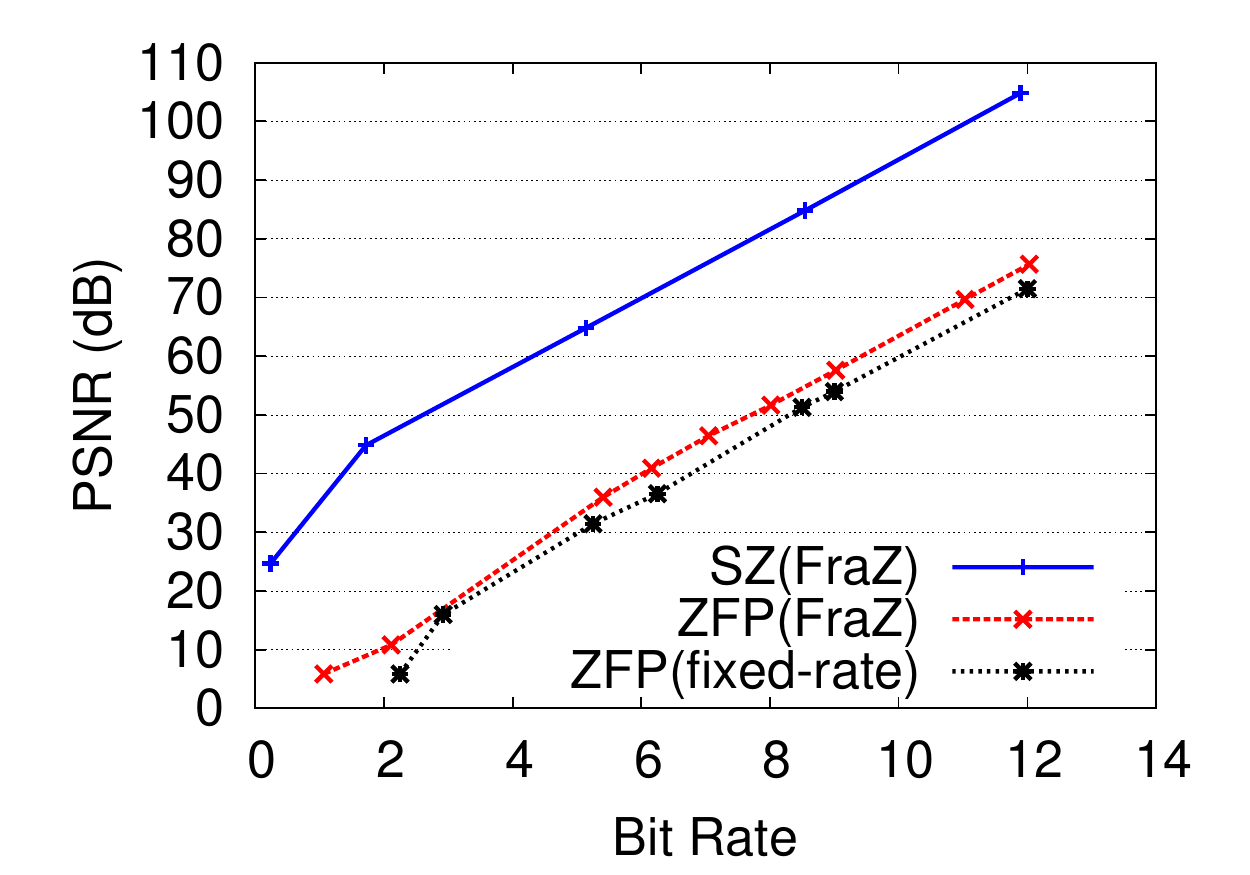}}
}
\hspace{-6mm}

\vspace{-2mm}
\caption{Rate distortion of lossy compression (MGARD is missing in (d) and (e) because it does not support 1D dataset)}
\label{fig:rate-distortion}
\end{figure}

Fields may take longer for a variety of reasons:
(1) the $\rho_t(D_{f,t})$ may not be feasible for one or more of the time-steps,
(2) the dataset may have higher entropy resulting in a longer encoding stage for algorithms such as SZ, or
(3) the fields may be of different sizes, and larger fields take longer.

\subsubsection{How does FRaZ compare with the existing fixed-rate compression methods in terms of rate distortion and visual quality?}
\label{sec:visualquality}



We present the rate distortion in Figure \ref{fig:rate-distortion}, which shows the bit rate (the number of bits used  per data point after the compression) versus the data distortion. Peak signal-to-noise ratio (PSNR) is a common indicator to assess the data distortion in the community. PSNR is defined as $20\cdot log_{10}(\frac{d_{max} - d_{min}}{rmse})$
where $rmse  = \sqrt{\frac{1}{N}\sum_{i=1}^{N} (d_i - d'_i)^2}$, and $d_{max}$ and $d_{min}$ refer to the max and min value, respectively. In general, the higher the PSNR, the higher the quality of decompressed data.

\begin{figure}[ht] \centering
\subfigure[{original raw data}]
{
\includegraphics[scale=.34]{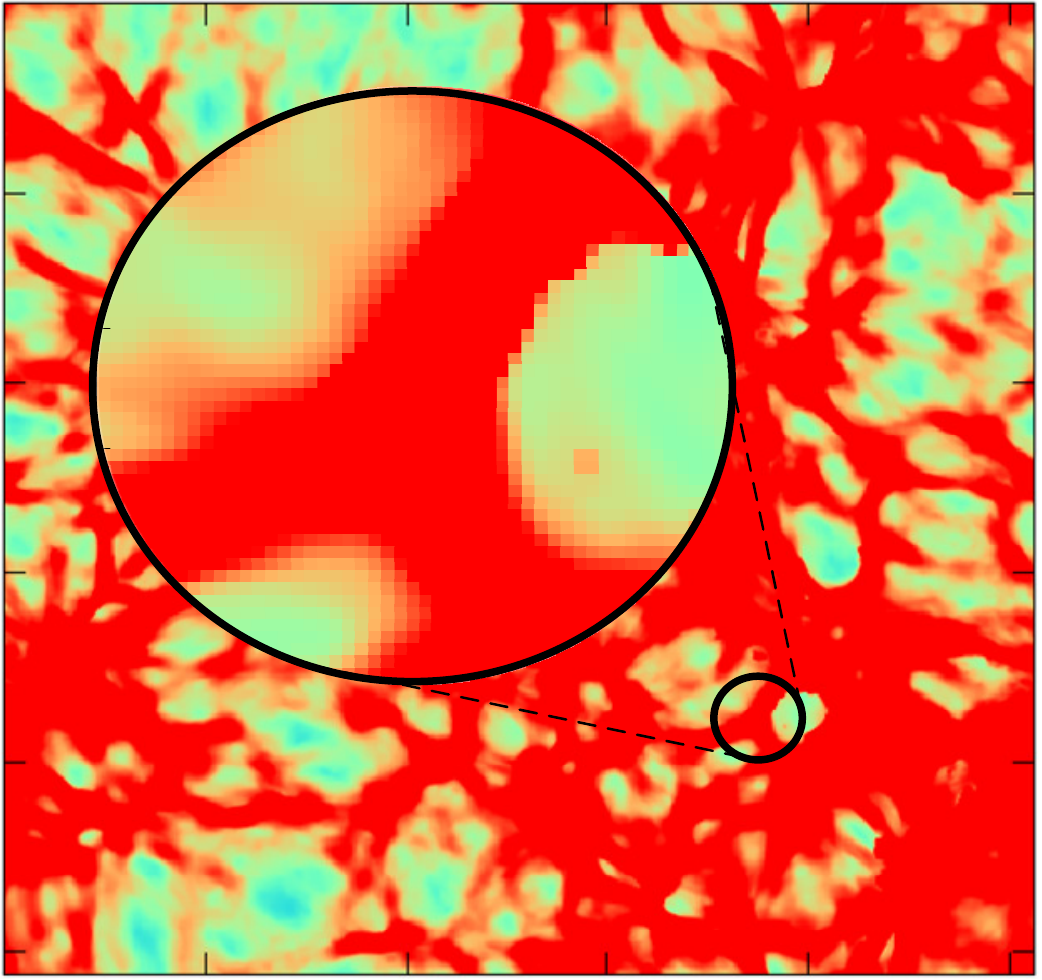}
}

\hspace{-3mm}
\subfigure[{ZFP (FRaZ) (PSNR=76, SSIM=0.997, ACF(error)=0.516)}]
{
\includegraphics[scale=.34]{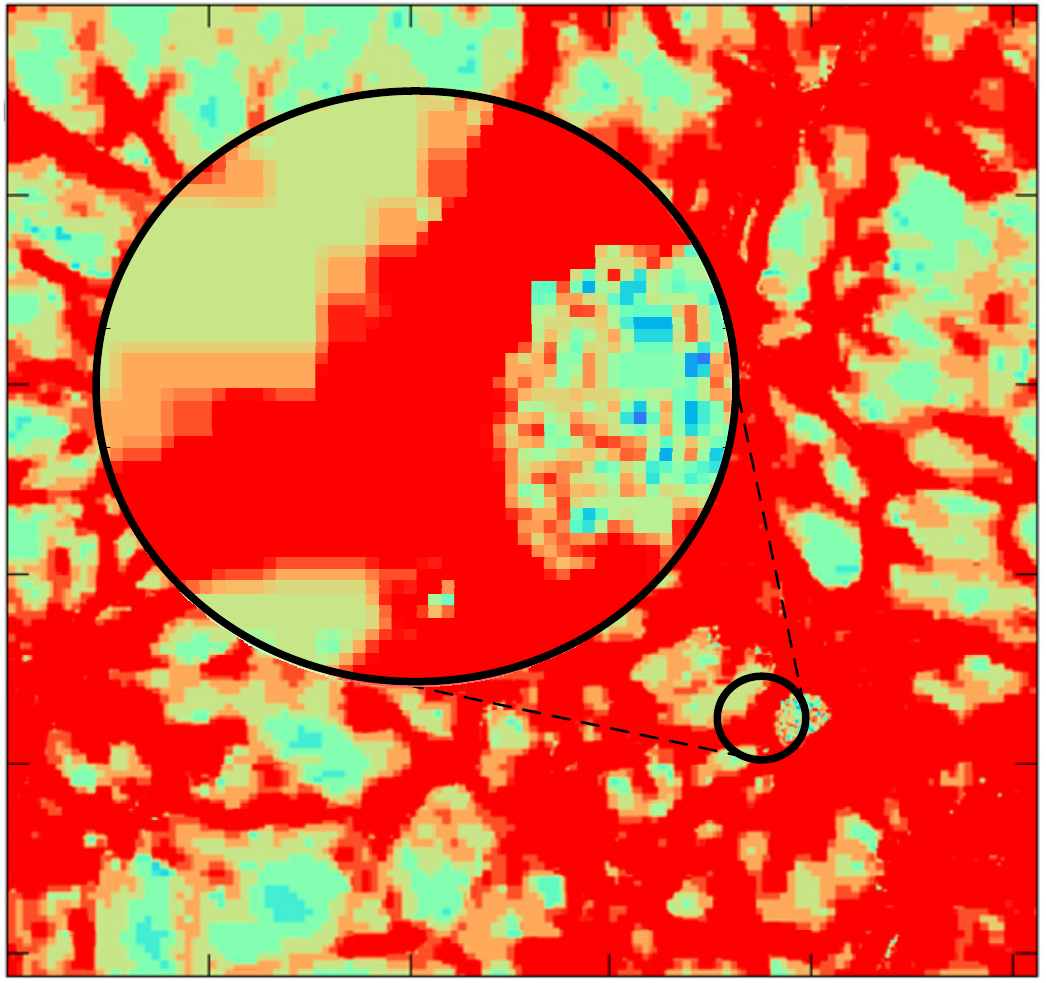}
}
\hspace{-3mm}
\subfigure[{ZFP (fixed-rate) (PSNR=56, SSIM=0.986, ACF(error)=0.383)}]
{
\includegraphics[scale=.34]{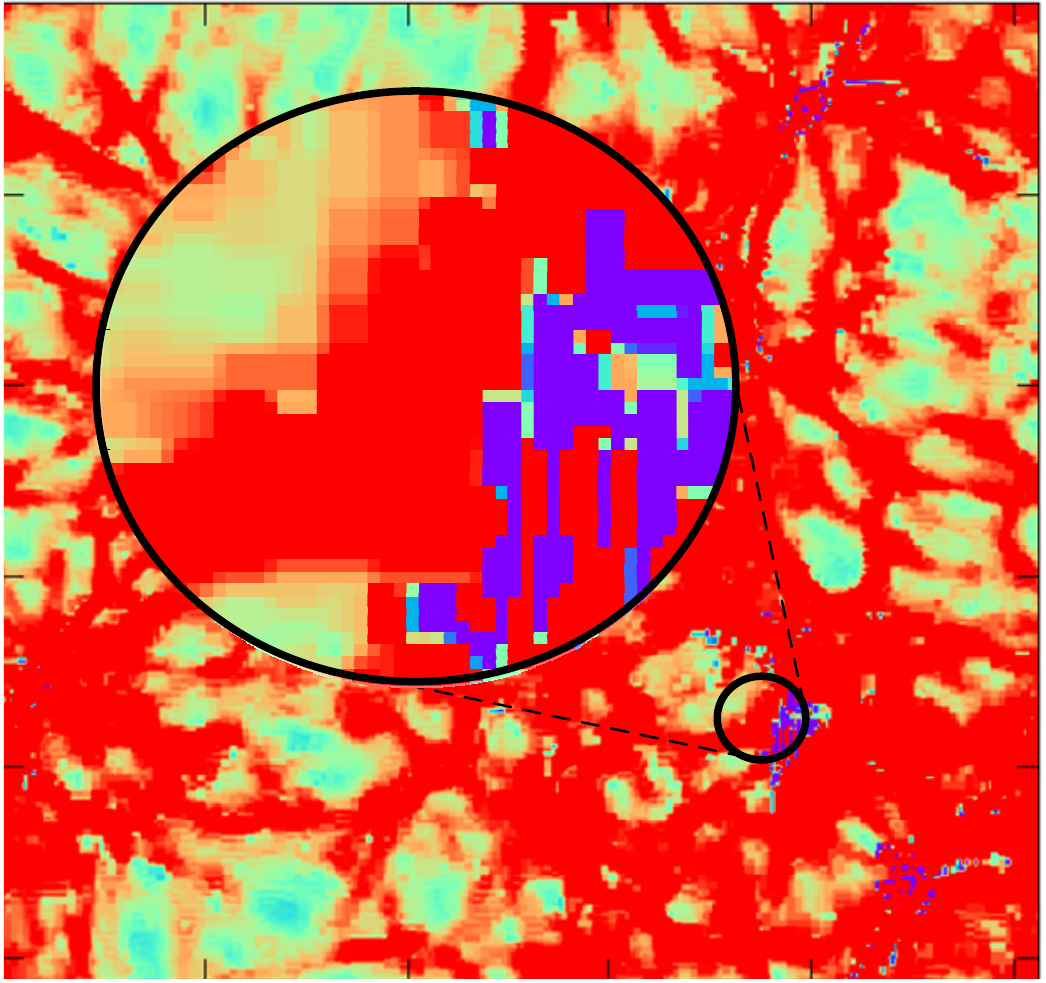}
}
\hspace{-6mm}

\hspace{-3mm}
\subfigure[{SZ (FraZ) (PSNR=80.4, SSIM=0.999, ACF(error)=0.344)}]
{
\includegraphics[scale=.34]{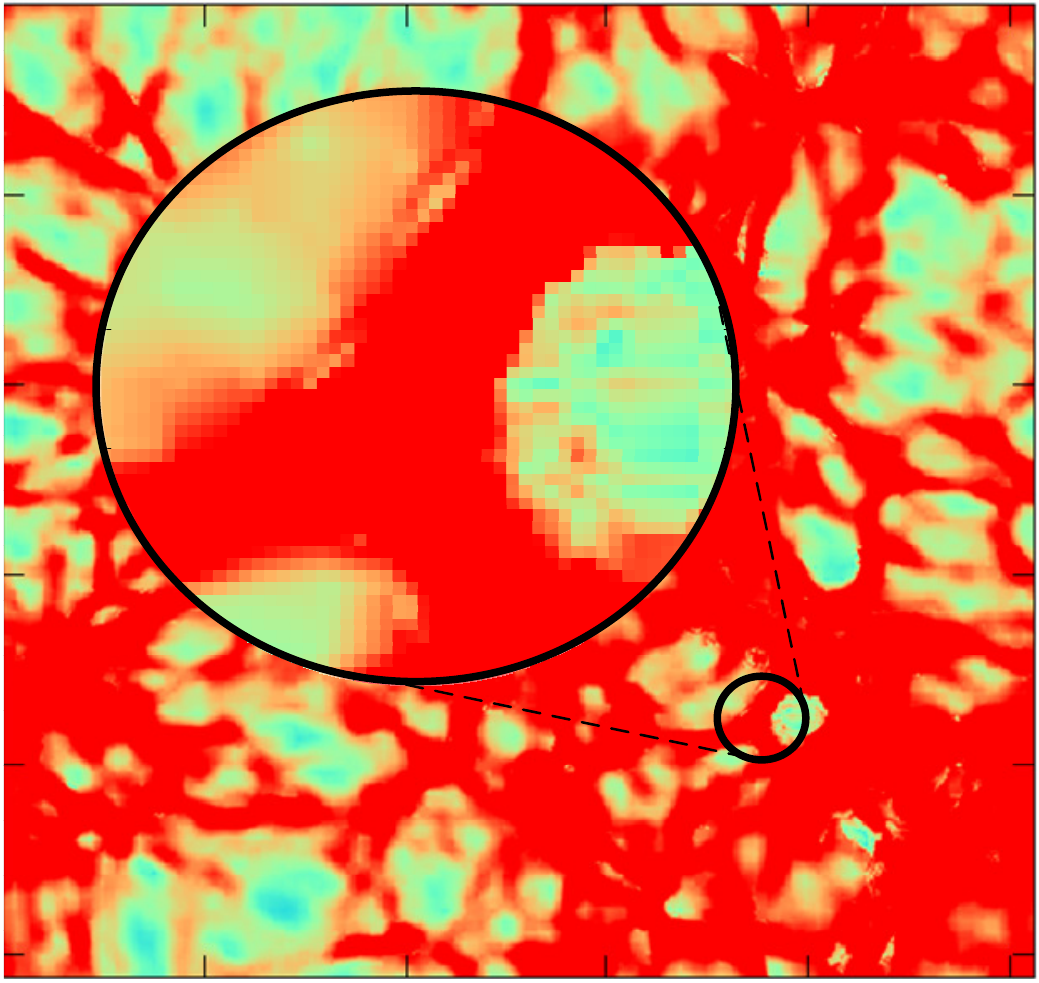}
}
\hspace{-3mm}
\subfigure[{MGARD (FraZ) (PSNR=70, SSIM=0.977, ACF(error)=0.92)}]
{
\includegraphics[scale=.34]{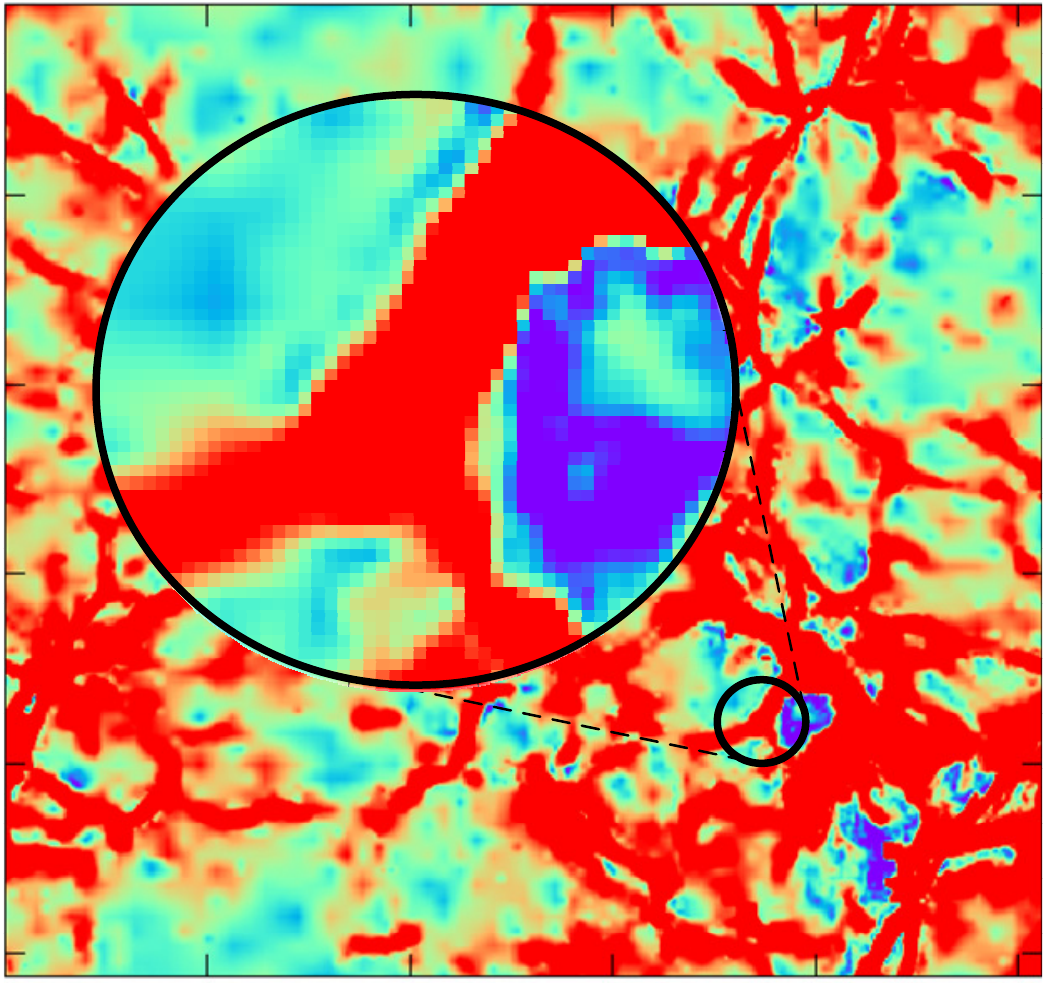}
}
\hspace{-6mm}

\vspace{-2mm}
\caption{Visualization of NYX (temperature:slice 256) with CR $\approx$ 85:1 (CR of ZFP(FRaZ) = 84, CR of ZFP(fixed-rate) = 85.3, CR of SZ(FRaZ) = 86, CR of MGARD(FRaZ) = 85.7)}
\label{fig:nyx-data-vis-temp}
\end{figure}

In this figure, one can clearly see that ZFP (FRaZ) provides consistently better rate distortion than does ZFP (fixed-rate) across bit-rates (i.e., across compression ratios). Moreover, SZ (FRaZ) exhibits the best rate distortion in most cases, which is consistent with the high compression quality  of SZ as presented in our prior work \cite{liangErrorControlledLossyCompression2018}. That being said, FRaZ can maintain the high fidelity of the data very well during the compression, by leveraging the error-bounded lossy compression mode for different compressors. 

In addition to the rate distortion, we present in Figure \ref{fig:nyx-data-vis-temp}  visualization images based on the same target compression ratio, to show fixed-ratio compression approach preserves visual quality. We wanted to set compression ratio of 100:1, but the closest fesible compression ratio for ZFP is $\sim$85:1 (see Section ~\ref{sec:exp-results}). Hence, we set the target compression ratio to be 85:1 for all compressors. Because of the space limit, we present the results only for NYX-temperature field;  other fields/applications exhibit similar  results. All the results are generated by FRaZ except for the ZFP(fixed-rate). ZFP(FRaZ) exhibits a much higher visual quality than does ZFP(fixed-rate) (see Figure \ref{fig:nyx-data-vis-temp} (b) vs. Figure \ref{fig:nyx-data-vis-temp} (c)), because FRaZ tunes the error bound based on fixed-accuracy mode, which has a higher compression quality than ZFP's built-in fixed-rate mode. ZFP(FraZ) exhibits higher PSNR than does ZFP(fixed-rate), which means higher visual quality. We also present the structural similarity index (SSIM)\cite{ssim-wang-2004} for the slice images shown in the figure. SSIM indicates similarity in luminance, contrast, and structure between two images; the higher SSIM, the better. Our evaluation shows that ZFP(fixed-rate) has lower SSIM than ZFP(FRaZ) -- i.e. better quality. From among all the compressors here, MGARD(FRaZ) leads to the lowest visual quality (as well as lowest PSNR and SSIM), because of inferior compression quality of MGARD on this dataset. 
\vspace{-2mm}
\section{Conclusions and Future Work}
\vspace{-1mm}
\label{sec:conclusion}

We have presented a functional, parallel, black-box autotuning framework that can produce fixed-ratio error-controlled lossy compression for scientific floating-point HPC datasets.
Our work offers improvements over existing fixed-rate methods by better preserving the data quality for equivalent compression ratios.
We showed that FRaZ works well for a variety of datasets and compressors.
We discovered that FRaZ generally has lower runtime for dataset and compressor combinations that produce large numbers of feasible compression ratios.

A number of areas for potential improvement exist.
First, we would like to consider arbitrary user error bounds.
By user error bounds, we mean error bounds that correspond with the quality of a scientist's analysis result relative to that on noncompressed data, 
such as \cite{bakerEvaluatingImageQuality2019} which identifies a particular SSIM in lossy compressed data required for valid results in their field.
Second, we would like to develop an online version of this algorithm to provide in situ fixed-ratio compression for simulation and instrument data.
Third, we would like to further improve the convergence rate of our algorithm to make it applicable for more use cases.

\vspace{-2mm}
\section*{Acknowledgment}
\vspace{-1mm}
\footnotesize{This research was supported by the Exascale Computing Project (ECP), Project Number: 17-SC-20-SC, a collaborative effort of two DOE organizations - the Office of Science and the National Nuclear Security Administration, responsible for the planning and preparation of a capable exascale ecosystem, including software, applications, hardware, advanced system engineering and early testbed platforms, to support the nation's exascale computing imperative.

The material was supported by the U.S. Department of Energy, Office of Science, under contract DE-AC02-06CH11357, and supported by the National Science Foundation under Grant No. 1619253 and 1910197.

We acknowledge the computing resources provided on Bebop, which is operated by the Laboratory Computing Resource Center at Argonne National Laboratory.

This material is also based upon work supported by the U.S. Department of Energy, Office of Science, Office of Workforce Development for Teachers and Scientists, Office of Science Graduate Student Research (SCGSR) program. The SCGSR program is administered by the Oak Ridge Institute for Science and Education (ORISE) for the DOE. ORISE is managed by ORAU under contract number DE-SC0014664. All opinions expressed in this paper are the author’s and do not necessarily reflect the policies and views of DOE, ORAU, or ORISE.}

\bibliographystyle{IEEEtran}
\bibliography{paper}

\end{document}